\documentclass[sigconf]{acmart}
\usepackage{graphicx}
\usepackage{amsmath}
\usepackage{subcaption}
\usepackage{colortbl}
\def\equationautorefname~#1\null{Eq.~#1\null}
\def\figureautorefname~#1\null{Fig.~#1\null}
\def\tableautorefname~#1\null{Table~#1\null}
\def\sectionautorefname~#1\null{Section~#1\null}
\def\subsectionautorefname~#1\null{Section~#1\null}
\def\subsubsectionautorefname~#1\null{Section~#1\null}
\def\paragraphautorefname~#1\null{Paragraph~#1\null}
\def\subparagraphautorefname~#1\null{Subparagraph~#1\null}
\def\pageautorefname~#1\null{page~#1 \null}
\def\appendixautorefname~#1\null{Appendix~#1 \null}

\AtBeginDocument{%
  }

\copyrightyear{2026}
\acmYear{2026}
\setcopyright{cc}
\setcctype{by}
\acmConference[CHI '26]{Proceedings of the 2026 CHI Conference on Human Factors in Computing Systems}{April 13--17, 2026}{Barcelona, Spain}
\acmBooktitle{Proceedings of the 2026 CHI Conference on Human Factors in Computing Systems (CHI '26), April 13--17, 2026, Barcelona, Spain}
\acmDOI{10.1145/3772318.3791183}
\acmISBN{979-8-4007-2278-3/2026/04}

\newcommand{\HL}[1]{\textcolor{black}{#1}}

\newcommand{\HLtwo}[1]{\textcolor{black}{#1}}

\title[Skewed Dual Normal Distribution Model]{Skewed Dual Normal Distribution Model: Predicting \HLtwo{1D} Touch Pointing Success Rate for Targets Near Screen Edges}

\newcommand{\mt}{\mathit{MT}}

\newcommand{\sr}{\mathit{SR}}

\newcommand{\margin}{\textsc{Margin}}
\newcommand{\mae}{\mathit{MAE}}
\newcommand{\rmse}{\mathit{RMSE}}
\newcommand{\mape}{\mathit{MAPE}}

\begin{document}

\author{Nobuhito Kasahara}
\affiliation{%
  \institution{Meiji University}
  \streetaddress{4-21-1 Nakano}
  \city{Nakano-ku}
  \state{Tokyo}
  \country{Japan}
  \postcode{1648525}}
\authornote{This work was conducted when the first author was interning at LY Corporation.}
\email{cs242009@meiji.ac.jp}

\author{Shota Yamanaka}
\affiliation{%
  \institution{LY Corporation}
  \streetaddress{1-3 Kioicho}
  \city{Chiyoda-ku}
  \state{Tokyo}
  \country{Japan}
  \postcode{1028282}}
\email{syamanak@lycorp.co.jp}

\author{Homei Miyashita}
\affiliation{%
  \institution{Meiji University}
  \streetaddress{4-21-1 Nakano}
  \city{Nakano-ku}
  \state{Tokyo}
  \country{Japan}
  \postcode{1648525}}
\email{homei@homei.com}

\begin{abstract}
Typical success-rate prediction models for tapping exclude targets near screen edges; however, design constraints often force such placements. \HL{Additionally, in scrollable UIs any element can move close to an edge.} \HLtwo{In this work, we model how target--edge distance affects \HLtwo{1D} touch pointing accuracy.} We propose the Skewed Dual Normal Distribution Model, which assumes the tap coordinate distribution is skewed by a nearby edge. The results of \HL{two smartphone experiments} showed that, as targets approached the edge, the distribution's peak shifted toward the edge and its tail extended away. In contrast to prior reports, the success rate improved when the target touched the edge, suggesting a strategy of ``tapping the target together with the edge.'' By accounting for skew, our model predicts success rates across a wide range of conditions, including edge‑adjacent targets, thus extending coverage to the whole screen and informing UI design support tools.
\end{abstract}

\maketitle

\section{Introduction}
Modeling human motor performance and refining models are central topics in HCI.
Research on Fitts' law, a model predicting movement time (\(\mt\)) for pointing, is a typical example~\cite{fitts1954information}.
Another key usability indicator is the success rate (\(\sr\)), or conversely, the error rate, and predictive models have been studied.
A seminal line of work by Wobbrock et al.\ predicts error rates based on the speed–accuracy trade-off, i.e., errors increase when pointing faster~\cite{Wobbrock2008_ERModel,Wobbrock2011ER2D}.
Bi et al.\ showed that touch pointing success can be estimated using the \emph{Dual Gaussian Distribution Model}, which accounts for ambiguity in finger tap coordinates~\cite{Bi2016DualGaussian}.
Building on this model, many refinements tailored to diverse task conditions have been proposed.

Most of these models focus on targets in the center of the screen.
In touch pointing, $\mt$ increases and $\sr$ decreases as targets approach an edge, therefore, touch-based UIs should avoid placing UI elements near the screen edge~\cite{Avrahami2015EdgeTouch,Usuba2023EdgeTarget}.
Nevertheless, placing targets near the edge is common.
For example, a dense layout may force elements near the edge, or scrolling may move targets close to the edge.
Existing models have a limitation in such cases: they estimate $\sr$s uniformly, regardless of target–edge distance.
\HL{Considering that all UI elements can potentially approach the screen edge in scrollable UIs, existing models may only be applicable to the special case where targets are fixed at positions sufficiently far from all edges—top, bottom, left, and right.}

If we could accurately predict $\sr$s near the edge, UI designers could utilize the entire screen more efficiently while maintaining user accuracy.
For example, combined with tools that compute $\sr$s from on-screen element size~\cite{usuba24arxivTappy,Yamanaka24arXivFigmaTappy,LIFULL24tapAnalyzer}, designers could account for proximity to the edge.
\HL{Furthermore, it is conceivable that a UI enabling more accurate selection could be realized by dynamically adjusting its layout based on target–edge distance after each scrolling operation.}
Nonetheless, no model has been proposed that describes how $\sr$ changes as a function of target-edge distance.

\begin{figure}[t]
\centering
    \includegraphics[height = 5cm]{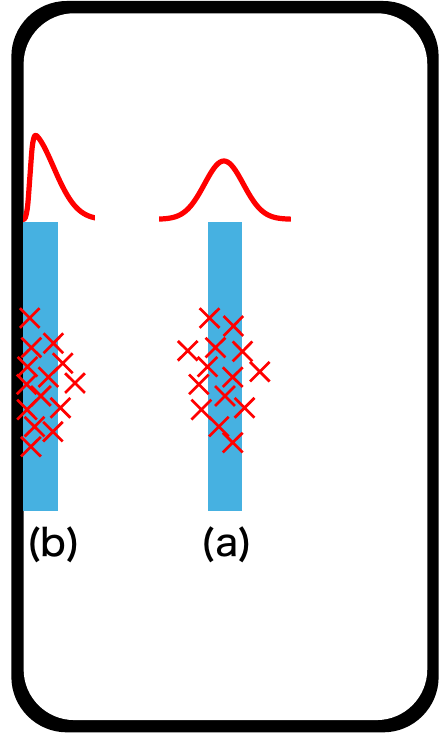}
\caption{The proposed Skewed Dual Normal Distribution Model assumes that the tap coordinate distribution is skewed by the presence of a screen edge on one side of the target and uses the cumulative distribution function of the skew-normal distribution to estimate tap success rate. (a) When the target is sufficiently far from the screen edge, the tap coordinate distribution is normal (Gaussian). (b) When the target is near the screen edge, the tap coordinate distribution becomes skew-normal.}
\label{fig:Fig1}
\Description{Two-panel schematic of the Skewed Dual Normal Distribution Model. (a) When a rectangular target is far from the screen edge, tap coordinates follow a symmetric normal (Gaussian) distribution centered on the target. (b) Near the edge, the tap-coordinate distribution becomes skew-normal: the peak shifts toward the edge and the tail extends away. The model estimates tap success by integrating the skew-normal CDF over the target bounds.}
\end{figure}

\HL{In light of this background, the purpose of this study is to model the relationship between target–edge distance and $\sr$.
Such interpretable mathematical models provide UI designers with quantitative and easy-to-understand design guidelines, and are considered useful for improving UI designs.}
We propose the Skewed Dual Normal Distribution Model, extending the Dual Gaussian Distribution Model~\cite{Bi2016DualGaussian} (\autoref{fig:Fig1}). 
Unlike existing models that assume a Gaussian distribution for tap coordinates, ours assumes that an edge near one side of the target skews the tap-point distribution and estimates $\sr$ using the skew-normal cumulative distribution function. 

We conducted \HL{two} 1D tap-pointing experiments.
\HL{One involved an edge present in the horizontal direction relative to the target, while the other involved an edge present in the vertical direction.}
\HL{While 2D targets are common in general UIs, our primary purpose was to investigate the effect of the distance from the screen edge on $\sr$.
Therefore, to isolate this effect by excluding other factors as much as possible, we decided to conduct 1D target experiments in two directions.}
\HL{\HLtwo{Through these experiments, we confirmed the model's applicability regardless of the axis orientation.}}
Unlike prior reports~\cite{Avrahami2015EdgeTouch,Usuba2023EdgeTarget,Henze2011LargeExperiment}, $\sr$ increased when the target touched the edge.
We attribute this to a strategy in which users ``tap the target together with the edge,'' avoiding errors on the side opposite the screen edge.

Our main contributions are as follows:
\begin{itemize}
  \item We theoretically derive an \HL{interpretable and highly accurate $\sr$ prediction model} that covers edge-adjacent targets previously treated as exceptions in $\sr$ modeling.
  \item We experimentally show that our model accurately predicts $\sr$s \HL{in both vertical and horizontal directions}, including targets near the edge, thereby extending the applicability of $\sr$ prediction models.
  \item We provide new insight into the ``tapping together with the edge'' user strategy as a factor underlying the proposed model's predictive accuracy.
\end{itemize}

\section{Related Work}
\subsection{Success Rate Estimation Models}
In HCI, $\sr$ prediction models for pointing have been widely studied.
Meyer et al.\ and Wobbrock et al.\ proposed $\sr$ prediction models in cursor-based pointing~\cite{Meyer1988Optimality,Wobbrock2008_ERModel,Wobbrock2011ER2D}.
In touch pointing, the model by Bi et al.\ is foundational~\cite{Bi2016DualGaussian}.
It assumes that tap positions follow a Gaussian distribution characterized by a mixture of variance due to target size and finger-placement uncertainty.
The tap-coordinate variances along the \(x\)- and \(y\)-axes can be estimated as a function of target width \(W\) and height \(H\) by
\begin{equation}
\sigma_{x}^2=a_{x}W^2+\sigma_{ax}^2\ \ \mathrm{and}\ \ \sigma_{y}^2=a_{y}H^2+\sigma_{ay}^2,
\label{Formula:Bi_sigma}
\end{equation}
where \(\sigma_x\) and \(\sigma_y\) are the standard deviations of tap coordinates along \(x\) and \(y\), and \(a_x,a_y,\sigma_{ax},\sigma_{ay}\) are the regression constants.
Using \(\sigma_x\) and \(\sigma_y\) estimated from target size, the probability that a tap falls inside a 2D target region \(D\) (i.e., \(\sr\)) is
\begin{equation}
\sr=\iint_D \frac{1}{2 \pi \sigma_{x}\sigma_{y}}\mathrm{exp}\left(-\frac{x^2}{2\sigma_{x}^2}-\frac{y^2}{2\sigma_{y}^2}\right) dxdy.
\label{Formula:DualGaussian}
\end{equation}
For 1D targets, this is simplified into~\autoref{Formula:DualGaussian1D}:
\begin{equation}
\sr = \text{erf}\left(\frac{W}{2\sqrt{2}\sigma}\right),
\label{Formula:DualGaussian1D}
\end{equation}
where \(\sigma\) is the tap-coordinate standard deviation along the target width direction, obtained using \autoref{Formula:Bi_sigma}.
These models have been shown to predict $\sr$s accurately~\cite{Bi2016DualGaussian,Yamanaka2020Rethinking,Yamanaka2021Crowdsource}.

Building on the Dual Gaussian Distribution Model, numerous \(\sr\) models have been proposed: 1D mouse pointing~\cite{Yamanaka2021Crowdsource}, moving targets~\cite{Huang2018Moving1D,Park2018,Lee18cue,Huang2019MovingTarget2D}, arbitrary target shapes~\cite{Zhang2020ArbitraryShapedMovingTarget,Zhang2023ArbitraryShape}, VR~\cite{Yu2019ERModelVR}, on-screen start tasks~\cite{Yamanaka2020Rethinking}, different finger travel directions~\cite{Ma2021RotationalDualGaussian}, and latency~\cite{Yu2023TemporalERSpatialCorrespondence}.
We aim to refine the Dual Gaussian Distribution Model to capture pointing accuracy near the edge.

\HL{Beyond these $\sr$ models, there is also a line of work that uses simulation models to reproduce full pointing trajectories, which can take several days for machine learning~\cite{Do2021SimulationModel,Fischer2022FeedbackControl}.
While these models are powerful tools for studying motor control, we seek a compact analytical model that predicts the average $\sr$ from UI layout parameters.
}

\subsection{Pointing Near Screen Edges}


For finger-based touch pointing, edge targets have been suggested to degrade performance.
Avrahami and Usuba et al. showed that smaller target–edge gaps increase $\mt$s~\cite{Avrahami2015EdgeTouch,Usuba2023EdgeTarget}.
A large-scale study (over 120 million taps) using an Android game app reported that targets near the edges are less accurate~\cite{Henze2011LargeExperiment}.
When nearby distractors exist, $\mt$ and $\sr$ worsen, and click distributions deviate from the normal distribution~\cite{Yamanaka2018PenalDistractors,Yamanaka2018SurroundingDistractors,Yamanaka2019DistractorCrowdSource}.


Prior work has offered quantitative guidelines (e.g., distancing targets \(\geq\) 4 mm from edges/distractors)~\cite{Avrahami2015EdgeTouch,Usuba2023EdgeTarget,Yamanaka2018PenalDistractors}.
However, we found no studies that model how target–edge distance affects pointing accuracy.
Such a model would enable UI designs that also exploit edge-adjacent areas efficiently.

\section{Skewed Dual Normal Distribution Model}
\subsection{Model Overview}

\begin{figure*}[t]
\centering
\begin{minipage}[b]{0.21\textwidth}
    \centering
    \includegraphics[height = 3.8cm]{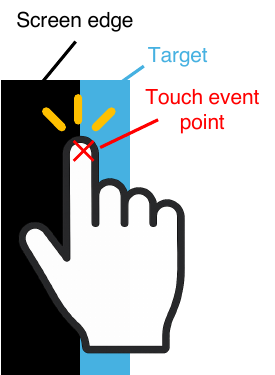}
    \subcaption{Touch event near edge}
    \label{fig:FingerEdge}
\end{minipage}
\begin{minipage}[b]{0.53\textwidth}
    \centering
    \includegraphics[height = 3.8cm]{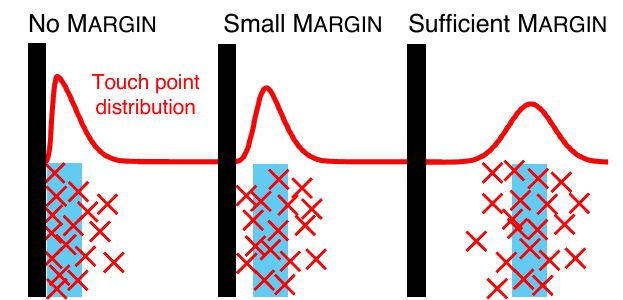}
    \subcaption{Distributional changes with edge distance}
    \label{fig:DistributionChange}
\end{minipage}
\begin{minipage}[b]{0.23\textwidth}
    \centering
    \includegraphics[height = 3.8cm]{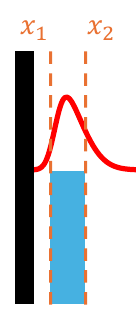}
    \subcaption{Computing \(\sr\) using the CDF}
    \label{fig:CalculateSR by CDF}
\end{minipage}
\caption{In the Skewed Dual Normal Distribution Model, (a) we assume that a touch that would have occurred outside the screen edge is triggered inside the screen due to finger thickness; (b) thus, as the target approaches the edge, the tap coordinate distribution becomes more skewed; and (c) we compute \(\sr\) using the skew-normal cumulative distribution function (CDF).}
\Description{Three-panel illustration of edge effects and success-rate computation. (a) A finger taps near the left screen edge; touches that would occur outside the display are registered just inside the edge. (b) As the gap ($\margin$) between target and edge shrinks from ‘Sufficient’ to ‘Small’ to ‘No’ margin, the tap distribution becomes increasingly skewed. (c) Success rate ($\sr$) equals the shaded probability mass between target bounds x1 and x2 under the skew‑normal curve (CDF).}
\label{fig:proposedModel}
\end{figure*}

While the Dual Gaussian Distribution Model assumes normally distributed taps, edge-adjacent targets may deviate from this assumption since taps cannot occur outside the edge.
Therefore, our model assumes a skew-normal distribution.
As illustrated in \autoref{fig:FingerEdge}, we posit that taps that would have occurred outside the edge are triggered inside due to finger thickness.
Consequently, the distribution may become more skewed the closer the target is to the edge (\autoref{fig:DistributionChange}).
The $\sr$ is then predicted from the skew-normal CDF (\autoref{fig:CalculateSR by CDF}).

Our assumption implies that more taps occur toward the edge, which conflicts with prior results suggesting that users avoid the edge~\cite{Henze2011LargeExperiment,Avrahami2015EdgeTouch,Usuba2023EdgeTarget}.
However, in Avrahami's experiment, a 3D-printed frame was placed on a large touchscreen~\cite{Avrahami2015EdgeTouch}, and the frame made it physically difficult to tap near the edge.
In addition, since circular targets touch the edge at a single point~\cite{Henze2011LargeExperiment,Usuba2023EdgeTarget}, \HL{users may have preferred the strategy of tapping away from the edge, rather than near it.}

In modern devices, bezels are flat or rounded toward the back, and our study used rectangular targets.
When the edge and target touch along a line, ``tapping together with the edge'' is a rational strategy to reduce errors on the side opposite the edge.
Indeed, our results show relatively higher \(\sr\)s for edge-adjacent targets, and open-ended comments indicated that participants adopted this strategy (discussed later).

\subsection{Probability of a Successful Tap}
For a 1D pointing task \HL{along either horizontal or vertical axis}, \(\sr\) is the probability mass of the tap coordinate distribution falling within the target bounds. 
The 1D skew-normal probability density function (PDF) is calculated as
\begin{equation}
f(x) = \frac{2}{\omega}\phi\left(\frac{x-\xi}{\omega}\right) \Phi\left(\alpha\frac{x-\xi}{\omega}\right),
\label{Formula:SkewNormalPDF}
\end{equation}
where \(\omega\) controls the scale, \(\xi\) controls the location, and \(\alpha\) controls the direction/degree of skew.
Here, \(\phi(z)\) and \(\Phi(z)\) are the PDF and CDF of the normal distribution, respectively:
\begin{gather}
\phi\left(z\right) = \frac{1}{\sqrt{2\pi}}\mathrm{exp}\left(-\frac{z^2}{2}\right) \label{Formula:NormalPDF} \\
\Phi\left(z\right) = \int_{-\infty}^z\phi(t)dt.\label{Formula:NormalCDF}
\end{gather}
For a target with bounds \(x_1 \le x_2\), the tap \(\sr\) is calculated as
\begin{equation}
\begin{split}
\sr &= P\left(x_1 \leq X \leq x_2\right) = \int_{x_1}^{x_2}f(x)dx \\
&= \int_{-\infty}^{x_2}f(x)dx - \int_{-\infty}^{x_1}f(x)dx,
\end{split}
\label{Formula:x1tox2}
\end{equation}
and the skew-normal CDF \(\int_{-\infty}^{x}f(t)dt\) is given using Owen's \(T\) function~\cite{Azzalini1985SkewNormal}:
\begin{equation}
\int_{-\infty}^{x}f(t)dt = \Phi\left( \frac{x-\xi}{\omega}\right)-2T\left(\frac{x-\xi}{\omega}, \alpha\right),
\label{Formula:SkewNormalCDF}
\end{equation}
with
\begin{equation}
T(h_T, a_T) = \frac{1}{2\pi}\int_{0}^{a_T}\frac{\mathrm{exp}\left(-\frac{1}{2}h_T^2\left(1+t^2\right)\right)}{1+t^2}dt,
\label{Formula:OwenT}
\end{equation}
where $h_T$ and $a_T$ are the parameters of Owen's $T$ function.
Using the error function, we obtain
\begin{equation}
\Phi\left(z\right) = \frac{1}{2}\left(1+\text{erf}\left(\frac{z}{\sqrt{2}}\right)\right).
\label{Formula:PhiERF}
\end{equation}
Setting the target center to \(x=0\), the $\sr$ for \HL{target size \(S\)} is the probability that the tap lies in \HL{\([{-}S/2,\, S/2]\).
Substitute target width $W$ or height $H$ for $S$ depending on the constraint direction of the 1D target (\autoref{fig:ModelAndExperiment Parameters} and \ref{fig:ModelAndExperiment ParametersBottom}).}
\begin{align}
\begin{split}
\sr = &\left(\frac{1}{2}\left(1+\text{erf}\left(\frac{\frac{\HL{S}}{2}-\xi}{\sqrt{2}\omega}\right)\right)-2T\left(\frac{\frac{\HL{S}}{2}-\xi}{\omega}, \alpha\right)\right) \\
&- \left(\frac{1}{2}\left(1+\text{erf}\left(\frac{-\frac{\HL{S}}{2}-\xi}{\sqrt{2}\omega}\right)\right)-2T\left(\frac{-\frac{\HL{S}}{2}-\xi}{\omega}, \alpha\right)\right).
\end{split}
\label{Formula:DualSkewSR}
\end{align}
Since the error function and Owen's \(T\) function are supported in many programming languages, practitioners can readily implement the model.
We use \texttt{scipy.special}'s \texttt{erf} and \texttt{owens\_t} functions, which employ accurate approximation algorithms~\cite{Johnson2012faddeeva,Patefield2000OwenT}.

\subsection{Parameters of the Skew-Normal Distribution}
\label{sec:parameters}
\begin{figure*}[t]
\centering
\begin{minipage}[b]{0.24\textwidth}
    \centering
    \includegraphics[height = 6cm]{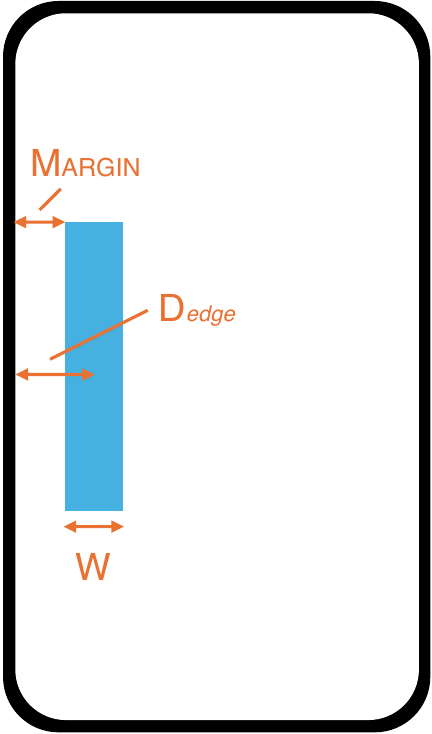}
    \subcaption{}
    \label{fig:ModelAndExperiment Parameters}
\end{minipage}
\begin{minipage}[b]{0.24\textwidth}
    \centering
    \includegraphics[height = 6cm]{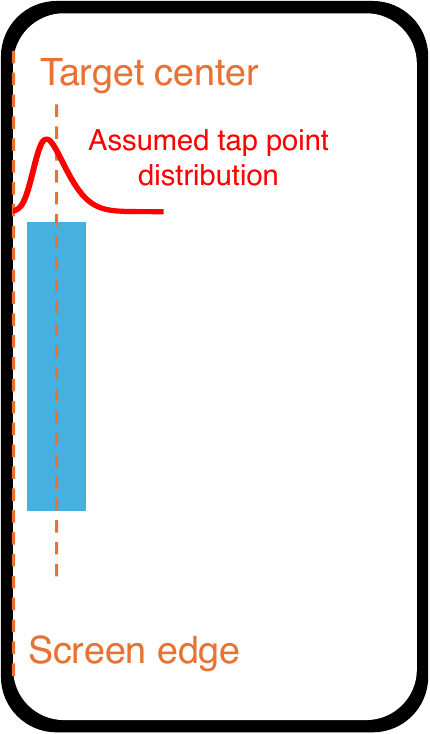}
    \subcaption{}
    \label{fig:EdgeAndTarget Left}
\end{minipage}
\begin{minipage}[b]{0.24\textwidth}
    \centering
    \includegraphics[height = 6cm]{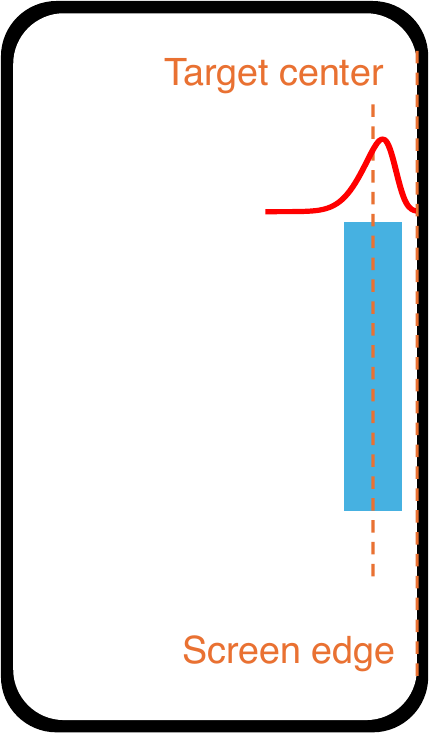}
    \subcaption{}
    \label{fig:EdgeAndTarget Right}
\end{minipage}
\begin{minipage}[b]{0.24\textwidth}
    \centering
    \includegraphics[height = 6cm]{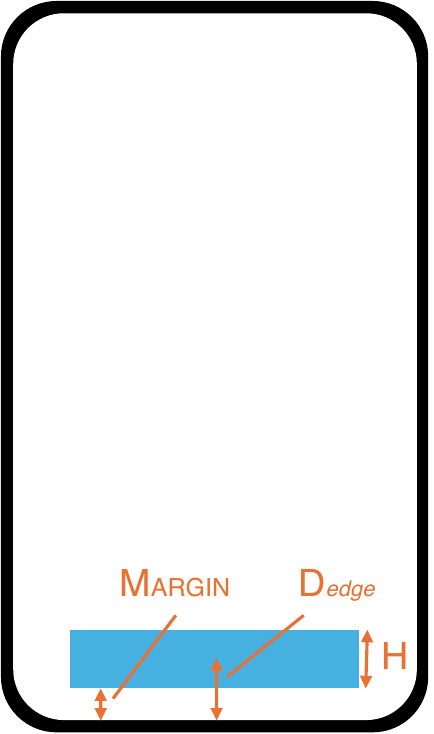}
    \subcaption{}
    \label{fig:ModelAndExperiment ParametersBottom}
\end{minipage}
\caption{Model parameters and experimental independent variables. (a) We denote the distance from the screen edge to the target center by \(D_\mathit{edge}\), the distance from the screen edge to the target edge by \(\margin\), and the target width by \(W\). We varied \(\margin\) and \(W\) in Experiment 1. (b) When the edge is on the left of the target, the distribution skews right (the peak shifts left); (c) when the edge is on the right, the distribution skews left (the peak shifts right). The sign of \(\gamma_1\) is set to \(\mathrm{sign}(\text{TargetCenter} - \text{ScreenEdge})\) to encode the skew direction (\autoref{Formula:DualSkewNormalEstimateGamma1}). \HL{(d) We varied vertical \(\margin\) and target height \(H\) in Experiment 2.}}
\Description{Model parameters and experimental independent variables. (a) Definitions: target width $W$, center‑to‑edge distance $D_{edge}$, and the edge‑to‑target gap $\margin$. (b) With the screen edge on the left, the predicted tap distribution is right‑skewed (peak shifts left, longer right tail). (c) With the edge on the right, the distribution mirrors (left‑skew). (d) In Experiment 2, we varied vertical $\margin$ and target height $H$ to test the model against the bottom screen edge.}
\end{figure*}

To predict \(\sr\) from \autoref{Formula:DualSkewSR}, we need \(\omega\) (scale), \(\xi\) (location), and \(\alpha\) (skew).
While maximum-likelihood or method-of-moments estimation is possible, such estimates can be unstable when the underlying distribution is close to normal~\cite{Pewsey2000ProblemOfAzzalini}.
Since tap coordinates are typically near-normal~\cite{Bi2016DualGaussian}, instead of modeling \(\omega,\xi,\alpha\), we model variance \(\sigma\), mean \(\mu\), and skewness \(\gamma_1\), and convert them by
\begin{gather}
\delta = \mathrm{sign}(\gamma_1) \times \min\left(0.999,\sqrt{\frac{\pi}{2}\frac{|\gamma_1|^{2/3}}{|\gamma_1|^{2/3}+\left(\frac{4-\pi}{2}\right)^{2/3}}}\right) \label{Formula:Delta}\\
\alpha = \frac{\delta}{\sqrt{1-\delta^2}} \label{Formula:Alpha}\\
\omega = \frac{\sigma}{\sqrt{1-\frac{2\delta^2}{\pi}}} \label{Formula:Omega}\\
\xi = \mu - \omega \times \delta \times \sqrt{\frac{2}{\pi}}. \label{Formula:Xi}
\end{gather}
Since \(|\delta| \ge 1\) makes \(\alpha\) undefined, we cap \(|\delta|\) at \(0.999\) (approximates an upper bound).
In our experiments, only one condition exceeded this cap.
When \(\gamma_1=0\), we have \(\delta=0\), \(\alpha=0\), \(\omega=\sigma\), \(\xi=\mu\), i.e., the normal distribution case, recovering the Dual Gaussian Distribution Model.
Then, if we can predict \(\sigma\), \(\mu\), and \(\gamma_1\) from target size \HL{(\(W\) or \(H\))} and target–edge distance (edge to center: \(D_\mathit{edge}\), edge to edge: \(\margin\); \autoref{fig:ModelAndExperiment Parameters} \HL{and \ref{fig:ModelAndExperiment ParametersBottom}}), we can predict \(\sr\) based on the task conditions.

Since the degree of skew increases as the target center approaches the left edge (\autoref{fig:DistributionChange}), \(|\gamma_1|\) increases as \(D_\mathit{edge}\) decreases and approaches 0 as \(D_\mathit{edge}\) grows large.
The sign of \(\gamma_1\) determines the skew direction: a positive \(\gamma_1\) implies right-skew (the peak shifts to the left with a longer right tail), and a negative \(\gamma_1\) implies the reverse.
We encode the sign by the relative positions of the edge and the target (\autoref{fig:EdgeAndTarget Left} and \autoref{fig:EdgeAndTarget Right}):
\begin{equation}
\label{Formula:DualSkewNormalEstimateGamma1}
\gamma_1 = \mathrm{sign}(\text{Target Center} - \text{Screen Edge}) \times \max\left(0,\, c + d \times D_\mathit{edge}\right),
\end{equation}
with regression constants \(c>0\) and \(d<0\).

As tap events cannot occur outside the edge, \(\sigma\) should decrease as \(\margin\) shrinks, and when \(\margin\) is sufficiently large, the model should recover to \autoref{Formula:Bi_sigma}.
We therefore model \(\sigma\) as
\begin{equation}
\label{Formula:DualSkewNormalEstimateSigma}
\begin{split}
&D_\mathit{edge} < -\frac{c}{d}:\ \ \sigma^2 = e + f \times \HL{S}^2 + g \times \margin,\\
&D_\mathit{edge} \ge -\frac{c}{d}:\ \ \sigma^2 = h + i \times \HL{S}^2,
\end{split}
\end{equation}
with regression constants \(e,f,g,h,i\).
Here \(-\frac{c}{d}\) (from \autoref{Formula:DualSkewNormalEstimateGamma1}) is the smallest \(D_\mathit{edge}\) for which the model predicts \(\gamma_1=0\).
Thus, for \(D_\mathit{edge} < -{c}/{d}\), skew is nonzero and \(\margin\) affects \(\sigma\); otherwise \(\gamma_1=0\), and we use the Dual Gaussian form.

\begin{figure}[t]
\centering
    \includegraphics[width = \columnwidth]{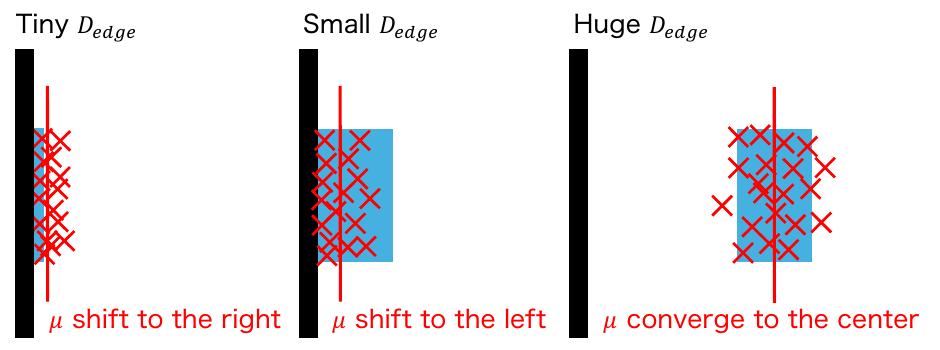}
\caption{\HL{Conceptual illustration of the modeling process for $\mu$. (Left) When $D_{edge}$ is very small, $\mu$ relative to the target center shifts in the positive direction. (Center) As the target size expands and $D_{edge}$ increases slightly, assuming that users adopt the strategy of ``tapping the target together with the edge,'' $\mu$ relative to the target center shifts in the negative direction. (Right) When $D_{edge}$ becomes sufficiently large and the influence of the screen edge vanishes, $\mu$ relative to the target center converges to the target center (zero).}}
\label{fig:ModelingMu}
\Description{Conceptual illustration of the modeling process for the mean offset $\mu$. (Left) When $D_{edge}$ is tiny (target touches edge), the mean shift is positive (away from edge) because off-screen taps are impossible. (Center) As $D_{edge}$ increases slightly, the mean shifts negative (toward the edge), reflecting a strategy of ‘tapping the target together with the edge.’ (Right) When $D_{edge}$ is huge, the mean converges to the target center (zero).}
\end{figure}

Existing models approximate $\mu = 0$ for simplification, i.e., the average tap position is close to the target center~\cite{Henze2011LargeExperiment}, even though it is influenced by various factors such as finger angle~\cite{Holz2011UnderstandingTouch}, on-screen target position~\cite{Azenkot2012TouchBehavior,Henze2011LargeExperiment}, grip posture~\cite{Lehmann2018HowToHold}, and target distance~\cite{Yu2019ERModelVR}.
\HL{Following this simplification, we approximated $\mu = 0$ when $D_{edge}$ is sufficiently large (i.e., $D_\mathit{edge} \ge -\frac{c}{d}$).}
\HL{For $D_\mathit{edge} < -\frac{c}{d}$, we modeled $\mu$ through the following steps (\autoref{fig:ModelingMu}).
First, we considered the extreme case where $D_{edge}$ approaches zero (i.e., a 1-pixel target touching the edge).
Since taps cannot be registered off-screen, the mean tap coordinate relative to the target center inevitably shifts away from the edge.
Next, as the target size increases while maintaining contact with the edge (increasing $D_{edge}$), we assumed that users adopt a strategy of ``tapping together with the edge'' to suppress errors on the opposite side.
Consequently, the mean coordinate relative to the target center shifts toward the edge.
Finally, when the target is sufficiently far from the edge, the mean should converge to the target center (zero).
Based on this non-monotonic transition—shifting away from the edge, then toward it, and finally returning to zero—we modeled $\mu$ using a quadratic function (\autoref{Formula:DualSkewNormalEstimateMu}).}
\begin{equation}
\label{Formula:DualSkewNormalEstimateMu}
\begin{split}
&\HL{D_\mathit{edge} < -\frac{c}{d}:\ \ \mu = j + k\left(D_\mathit{edge} - l\right)^2, }\\
&\HL{D_\mathit{edge} \ge -\frac{c}{d}:\ \  \mu = 0,}
\end{split}
\end{equation}
with regression constants \(j,k,\HL{l}\).

In summary, we estimate \(\gamma_1,\sigma,\mu\) from Eqs.~\ref{Formula:DualSkewNormalEstimateGamma1}–\ref{Formula:DualSkewNormalEstimateMu}, convert them into \(\delta,\alpha,\omega,\xi\) using Eqs.~\ref{Formula:Delta}–\ref{Formula:Xi}, and substitute them into \autoref{Formula:DualSkewSR} to predict the 1D \(\sr\).
\HL{The apparent mathematical complexity stems from parameter transformations and the CDF, but the predictive models require simple regressions.}
Extension to general 2D pointing is possible~\cite{Bi2016DualGaussian,Usuba2022ER1Dto2D}, \HL{but here we focus on 1D to isolate and examine the effect of a single screen edge.}

\section{Experiments}
\begin{figure}[t]
\centering
    \includegraphics[width = \columnwidth]{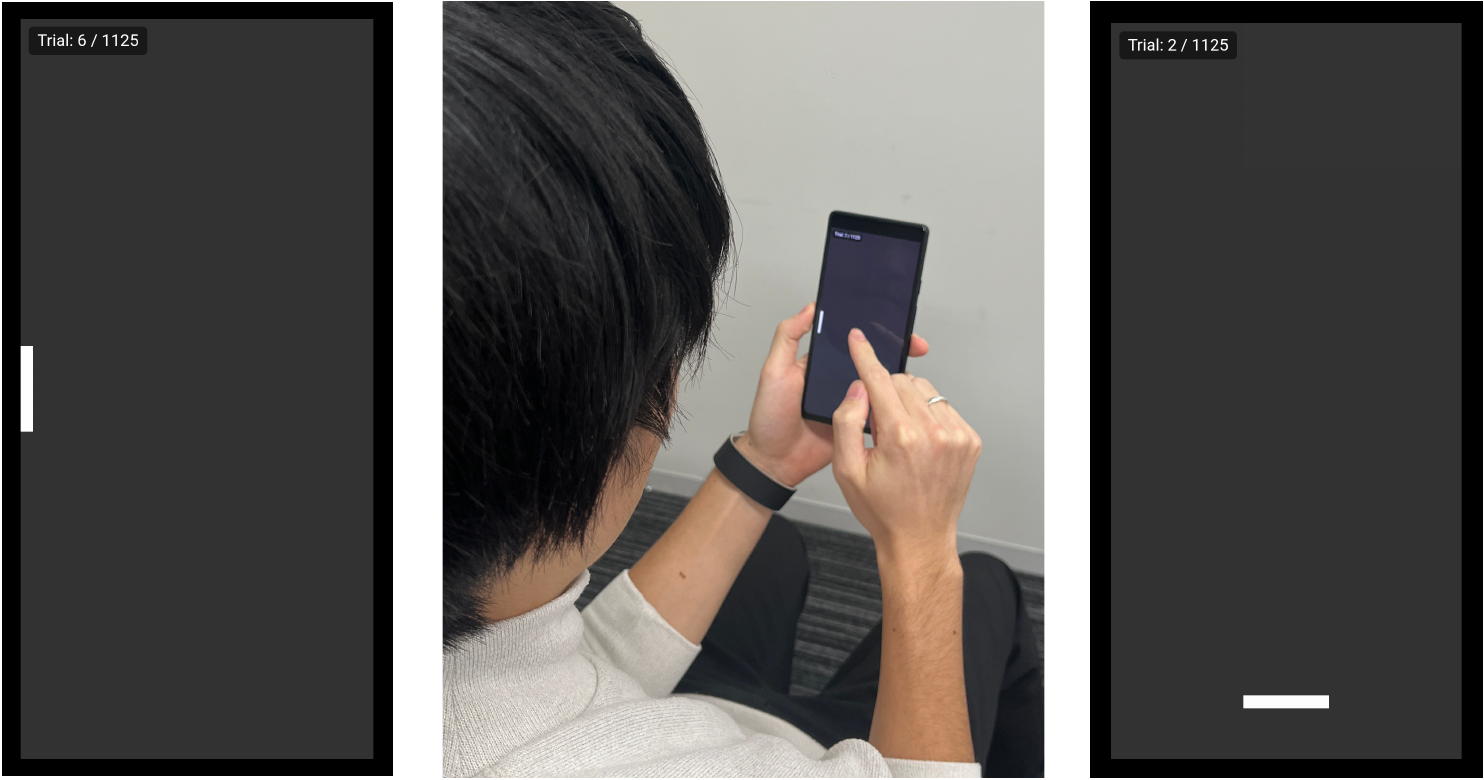}
\caption{(Left) Screenshot of the app used in Experimental 1 (HTML/CSS/JavaScript; displayed in Google Chrome fullscreen). (Center) Experimental setup. Participants sat, held the smartphone in the non-dominant hand, and performed the task with the dominant index finger. They returned their dominant hand to their knees between trials. \HL{This setup was consistent across Experiments 1 and 2.} \HL{(Right)  Screenshot of the app used in Experimental 2.}}
\label{fig:Experiment1}
\Description{Experimental setup. Left: app screenshot showing a tall white rectangular target aligned near the left edge on a dark smartphone screen. Right: participant seated, holding a Google Pixel 6a in the non‑dominant hand and tapping with the dominant index finger in an off‑screen start task; the hand returns to the knees between trials.}
\end{figure}

We conducted \HL{two} 1D tap pointing experiments to evaluate our model.
\HL{In Experiment 1, \(W\) and the distance from the left screen edge (\(\margin\)) are independent variables, and in Experiment 2, \(H\) and the distance from the bottom edge (\(\margin\)) are independent variables (\autoref{fig:Experiment1}).}
Participants held the smartphone in their non-dominant hand and tapped with their dominant index finger in an off-screen pointing task.
We excluded one-handed thumb input because \HL{individual differences in thumb length affect target reachability and \(\sr\)~\cite{Perry2008Thumb,BergstromLehtovirta14}. 
To isolate and validate the effect of a single edge, this study was limited to tapping with the dominant index finger.}
While on-screen tasks also conform to the Dual Gaussian~\cite{Yamanaka2020Rethinking}, we used an established off-screen task for a fair comparison with it~\cite{Bi2016DualGaussian}.

\subsection{\HL{Experiment 1: Left Screen Edge Task}}
\subsubsection{Participants}
Fifteen computer science students (4 female, 11 male; mean age 21.3, SD 1.73), all right-handed, participated in the experiment.
Compensation was 1{,}800 JPY (12.11 USD).
Before the experiment, participants received an explanation and consented to participate.
Research ethics review at the first author's institution was not mandatory for experiments involving these smartphone operation tasks.

\subsubsection{Apparatus}
We used a Google Pixel 6a (body: \(152.2\times 71.8\times 8.9\) mm; display: \(142.5\times 64.1\) mm; resolution: \(2400\times 1080\) px) without a case.
The Pixel 6a has a thin bezel curved toward the back, which should not hinder edge-adjacent tapping.
The system was implemented in HTML/CSS/JavaScript and shown in Google Chrome fullscreen.

\subsubsection{Design}
The within-subjects design was \(9\ \margin\text{s} \times 5\ W\text{s}\).
Targets were aligned relative to the left screen edge.
Factors included the distance from the left screen edge to the target's left edge (\(\margin\): 0, 1.560, 3.119, 4.679, 7.798, 9.358, 12.477, 15.596, 18.715~mm) and target width (\(W\): 1.560, 2.339, 3.119, 4.679, 7.798~mm)
\footnote{We used integer pixel values to avoid rounding errors and reported millimeter values with the necessary precision for reproducibility.\label{foot:ReasonMM}} (\autoref{fig:ModelAndExperiment Parameters}).
Since \(\margin\) is the gap between the screen edge and the target edge, \(D_\mathit{edge} = \margin + W/2\).
\HL{The target height was fixed at 15.596 mm in all conditions.}
\HL{Since prior research claimed that targets of 7.62 mm~\cite{Bi2016DualGaussian} or 9 mm~\cite{Yamanaka2020Rethinking} or larger achieve $\sr$s above 99\%, we set the height to bigger than those thresholds.}
We decided to use a large but finite height to eliminate the effects of the top and bottom screen edges.
\HL{For example, participants might adopt different operational strategies when tapping near the top edge, bottom edge, or center of the screen on an infinite height target.}
\HL{Our targets were sufficiently large that errors effectively did not occur in the height direction, allowing it to be regarded as a 1D target~\cite{Accot03BivariatePointing,fitts1954information}. 
Indeed, such errors were 29 out of 16,200 trials (0.179\%) in our Experiment 1.}
A \textit{set} consisted of the 45 target conditions presented in random order, and participants completed \(25\ \text{sets}\) (\(9\ \margin\text{s} \times 5\ W\text{s} \times 25\ \text{sets} = 1125\) trials).

We recorded tap coordinates and whether the trial was successful at finger-up~\cite{Bi2016DualGaussian,Yamanaka2020Rethinking}.
Dependent variables were \(\sigma\), \(\mu\), \(\gamma_1\), and \(\sr\) (the fraction of trials in which the first tap hit the target).

\subsubsection{Procedure}
Participants were instructed to tap the white target displayed on the screen ``as fast and as accurately as possible'' (\autoref{fig:Experiment1}).
On errors (tapping outside the target), audio feedback and a blinking yellow target were presented, and participants had to tap the target again.
On success, the success audio played, and the screen went dark for 0.5 s.
Participants returned their dominant hand to their knees during the blackout and then tapped the next target.
\HL{This \HLtwo{instruction} may have increased 3D hand movements compared to the instruction in prior research to place the hand ``off the screen in natural and comfortable positions''~\cite{Bi2016DualGaussian}.
However, \HLtwo{by enforcing a specific off-screen location, we strictly controlled the starting condition to ensure that no movements started from on-screen area.}}
Participants could take breaks at any time; additionally, a mandatory 30-second break followed every five sets. 
After completing all tasks, we collected data on gender, age, and open-ended comments about manipulation strategies.
All participants finished within 30–60 minutes.

\subsection{\HL{Experiment 2: Bottom Screen Edge Task}}
\subsubsection{\HL{Overview}}
\HL{The apparatus and procedure were identical to those in Experiment 1, while the participants and experimental design were changed.
No participants took part in Experiment 1.
The experimental design was modified to accommodate the shift from examining horizontal 1D tap $\sr$ to vertical 1D $\sr$.
}

\subsubsection{\HL{Participants}}
\HL{Fifteen computer science students (5 female, 10 male; mean age 20.9, SD 1.94; all right-handed) participated in Experiment 2.
The procedures for compensation, informed consent, and research ethics were identical to those in Experiment 1.
}

\subsubsection{\HL{Design}}
\HL{The within-subjects design was \(9\ \margin\text{s} \times 5\ H\text{s}\).
While targets were aligned relative to the bottom screen edge, we used the same \(\margin\)s and target sizes ($H$ instead of $W$) as in Experiment 1 (\autoref{fig:ModelAndExperiment ParametersBottom}).
\HL{The target width was fixed at 15.596 mm in all conditions.}
The errors recorded in the width direction were 58 out of 16,200 trials (0.358\%).
The number of sets, task presentation order, and dependent variables were identical to those of Experiment 1.
}

\section{\HL{Results of Experiment 1}}
\subsection{Outliers}
We discarded the first set as practice.
Of the remaining \(16{,}200\) trials, 29 trials (0.179\%) with errors above or below the target were excluded to focus on horizontal $\sr$.
Then, for each participant and condition, we removed taps \(>3\) SD from the mean (97 trials, 0.600\%), following prior work~\cite{SOUKOREFF2004751}.
The exclusion rate lies within the reported range \HL{in some previous works} (about 0.5–7\%)~\cite{Yamanaka2021Crowdsource,Yamanaka2024ISS,Komarov2013Crowdsourcing,Schwab2019PanZoom}.
The analyses below utilize the remaining 16,074 trials.

\subsection{Distributional Tests of Tap Coordinates}
\begin{figure}[t]
\centering
    \includegraphics[width = \columnwidth]{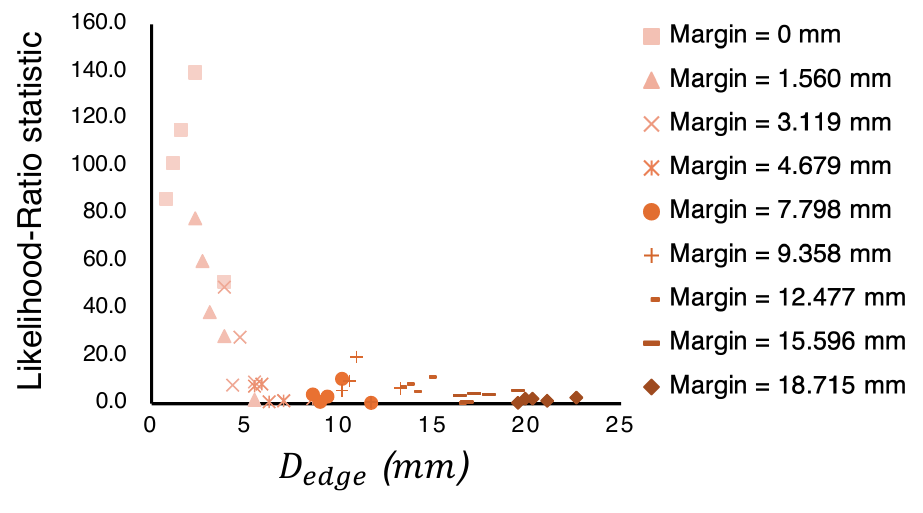}
\caption{\(D_\mathit{edge}\) vs.\ likelihood-ratio statistic in Experiment 1. Likelihood-ratio tests indicate that the closer the target is to the edge, the more the tap coordinate distribution follows a skew-normal distribution rather than a normal distribution.}
\Description{Scatter plot of $D_{edge}$ (mm) versus the likelihood‑ratio statistic in Experiment 1 (Left Edge). Points are grouped by Margin level. The statistic increases sharply as $D_{edge}$ decreases, indicating that the skew‑normal distribution fits the tap data significantly better than the normal distribution when targets are near the edge.}
\label{fig:Dedge vs LikelihoodRatio Left}
\end{figure}

Whereas the existing model assumes normality, our model assumes skew-normality.
Thus, we tested the distributional fit.
A Shapiro–Wilk test showed that four of the 45 conditions (8.89\%) were normally distributed, suggesting limits to the assumption of the existing model.
Likelihood-ratio tests comparing normal vs.\ skew-normal indicated that the statistic increases as \(D_\mathit{edge}\) decreases (\autoref{fig:Dedge vs LikelihoodRatio Left}).
Thus, proximity to the edge increases skew-normality, supporting our model's assumption.

\subsection{Prediction Accuracy of the Models}
\begin{table*}[t]
\caption{Regression constants and accuracy metrics for each model in Experiment 1. \HLtwo{The number of parameters in machine learning models refers the quantity of parameters necessary for making predictions.} \HL{Although the $R^2$ for the MLP Neural Net was lower than -1, which might appear unnatural, it is a possible outcome\protect\footnotemark. Since MLP Neural Nets strongly require hyperparameter tuning and feature scaling, significantly low $R^2$ values can occur under default settings\protect\footnotemark.}}
\centering
\begin{tabular}{c|c|c|c|c|c|c|c|c}
\multicolumn{2}{c|}{} & \multicolumn{5}{c}{Regression Analysis} & \multicolumn{2}{|c}{LOOCV}\\
\hline
Model & Equation & Regression Constants & $R^2$ & $\mae$ & $ \rmse$ & $\mape$ & $R^2$ & $\mae$\\
\hline
Gaussian & $\sigma^2$ (Eq.~\ref{Formula:Bi_sigma}) & $a = 1.50$, $b = 0.0236$ & $.437$ & $0.426$ & 0.574 & $37.6\%$ & $.390$ & $0.445$\\
\cline{2-9}
& $\sr$ (Eq.~\ref{Formula:DualGaussian1D}) & - & $.816$ &  $5.44$ & $7.75$ & $8.11\%$ & $.807$ & $5.56$\\
\hline
Skewed & $\gamma_1$ (Eq.~\ref{Formula:DualSkewNormalEstimateGamma1}) & $c = 1.09$, $d = -0.170$ & $.789$ & $0.123$ & $0.149$ & $192\%$ & $.761$ & $0.130$\\
\cline{2-9}
& $\sigma$ (Eq.~\ref{Formula:DualSkewNormalEstimateSigma}) & \begin{tabular}{c}$e = 0.155$, $f = 0.0461$,\\$g = 0.466$, $h = 1.60$,\\ $i = 0.0205$\end{tabular} & $.882$ & $0.0753$ & $0.102$ & $5.67\%$ & $.851$ & $0.0855$\\
\cline{2-9}
& $\mu$ (Eq.~\ref{Formula:DualSkewNormalEstimateMu}) & \begin{tabular}{c}\HL{$j = -0.393$, $k = 0.108$,}\\ \HL{$l = 3.73$}\end{tabular}  & \HL{$.905$} & \HL{$0.0752$} & \HL{$0.0895$} & \HL{$104\%$} & \HL{$.847$} & \HL{$0.0938$}\\
\cline{2-9}
& $\sr$ (Eq.~\ref{Formula:DualSkewSR}) & - & \HL{$.950$} & \HL{$3.23$} & \HL{$4.05$} & \HL{$4.85\%$} & \HL{$.944$} & \HL{$3.39$}\\
\hline
\multicolumn{9}{c}{\HL{Machine Learning Models with Default Hyperparameters}}\\
\hline
\HL{Lasso Regression} & \HL{-} & \HLtwo{number of parameters: 3} & \HL{.743} & \HL{7.36} & \HL{9.17} & \HL{11.3\%} & \HL{.706} & \HL{7.88} \\
\hline
\HL{Random Forest} & \HL{-} & \HLtwo{number of parameters: 5,708} & \HL{.987} & \HL{1.48} & \HL{2.09} & \HL{2.24\%} & \HL{.903} & \HL{3.94} \\
\hline
\HL{SVR} & \HL{-} & \HLtwo{number of parameters: 45} & \HL{.213} & \HL{13.8} & \HL{16.0} & \HL{21.9\%} & \HL{.137} & \HL{14.6} \\
\hline
\HL{MLP Neural Net} & \HL{-} & \HLtwo{number of parameters: 401} & \HL{-2.65} & \HL{28.2} & \HL{34.5} & \HL{38.2\%} & \HL{-2.64} & \HL{28.4} \\
\hline
\multicolumn{9}{c}{\HL{Machine Learning Models with Tuned Hyperparameters}}\\
\hline
\HL{Lasso Regression} & \HL{-} & \HLtwo{number of parameters: 3} & \HL{.714} & \HL{7.35} & \HL{9.67} & \HL{11.9\%} & \HL{.693} & \HL{7.61} \\
\hline
\HL{Random Forest} & \HL{-} & \HLtwo{number of parameters: 8,445} & \HL{.925} & \HL{3.67} & \HL{4.95} & \HL{5.71\%} & \HL{.857} & \HL{4.74} \\
\hline
\HL{SVR} & \HL{-} & \HLtwo{number of parameters: 44} & \HL{.993} & \HL{0.652} & \HL{1.54} & \HL{1.07\%} & \HL{.968} & \HL{2.42} \\
\hline
\HL{MLP Neural Net} & \HL{-} & \HLtwo{number of parameters: 7,999} & \HL{.999} & \HL{0.230} & \HL{0.325} & \HL{0.349\%} & \HL{.966} & \HL{2.62} \\
\hline
\end{tabular}
\label{table:Model Regression Ex1}
\end{table*}

\subsubsection{Dual Gaussian Distribution Model}
\begin{figure*}[t]
\centering
\begin{minipage}[b]{0.45\textwidth}
    \centering
    \includegraphics[width=0.95\columnwidth]{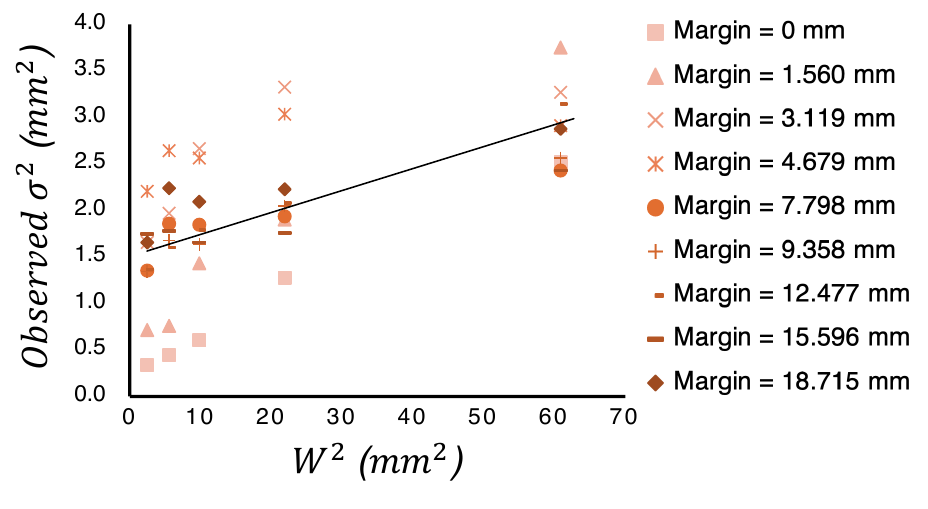}
    \subcaption{$W^2$ vs.\ $\sigma ^2$}
    \label{fig:DualGaussian Sigma}
\end{minipage}
\begin{minipage}[b]{0.45\textwidth}
    \centering
    \includegraphics[width=0.95\columnwidth]{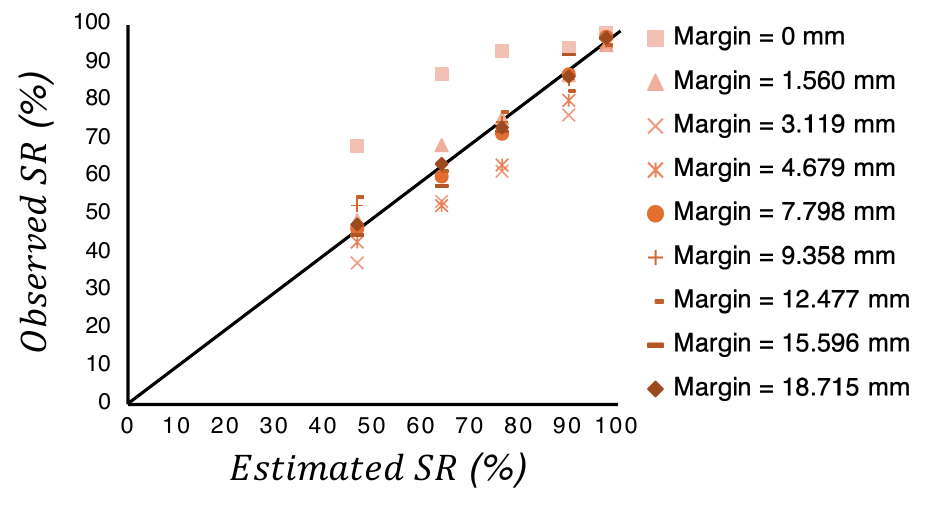}
    \subcaption{Predicted \(\sr\) vs.\ Observed \(\sr\)}
    \label{fig:DualGaussian SR}
\end{minipage}
\caption{Results for the Dual Gaussian Distribution Model in Experiment 1. Large deviations appear when \(\margin\) is small, indicating the need to model edge distance in \(\sigma\) and \(\sr\). In (a), the line is the regression line; in (b), the line indicates perfect prediction.}
\Description{Dual Gaussian model diagnostics for Experiment 1. (a) $W^2$ vs observed $\sigma^2$ with a regression line: deviations grow at small $\margin$s, indicating unmodeled edge effects on spread ($R^2 = .437$). (b) Predicted $\sr$ vs observed $\sr$ with a 45° reference line: under‑prediction is visible for near‑edge conditions.}
\end{figure*}

We regressed \(\sigma\) \HL{of the horizontal tap coordinates} (averaged over participants per condition) using \autoref{Formula:Bi_sigma}. 
Fit was modest (\(R^2=.437\); \autoref{fig:DualGaussian Sigma}, \autoref{table:Model Regression Ex1}), with larger deviations at small \(\margin\), suggesting that a width-only model fails to capture edge-induced changes in \(\sigma\).

Using the estimated \(\sigma\) in \autoref{Formula:DualGaussian1D}, we predicted \(\sr\) and observed unadjusted \(R^2=.816\) (\autoref{fig:DualGaussian SR}, \autoref{table:Model Regression Ex1}).
Especially at small \(\margin\), the observed \(\sr\) exceeded predictions.
Possible causes include the placeholder effect~\cite{Adam2006MovingFarther,Bradi2009ModulatingFitts,Pratt2007VisualLayout} and the ``tap together with the edge'' strategy, both of which improve accuracy near the edge.

\footnotetext[2]{\HL{\url{https://scikit-learn.org/stable/modules/generated/sklearn.metrics.r2_score.html}}}
\footnotetext[3]{\HL{\url{https://scikit-learn.org/stable/modules/neural_networks_supervised.html}}}

We assessed generalization by performing leave-one-out cross-validation (LOOCV) over task conditions (\autoref{table:Model Regression Ex1}).
LOOCV \(R^2\) and \(\mae\) closely matched the in-sample results, indicating no overfitting and similar predictive power under untested conditions.

\subsubsection{Proposed Model}
\begin{figure*}[t]
\centering
\begin{minipage}[b]{0.45\textwidth}
    \centering
    \includegraphics[width=0.95\columnwidth]{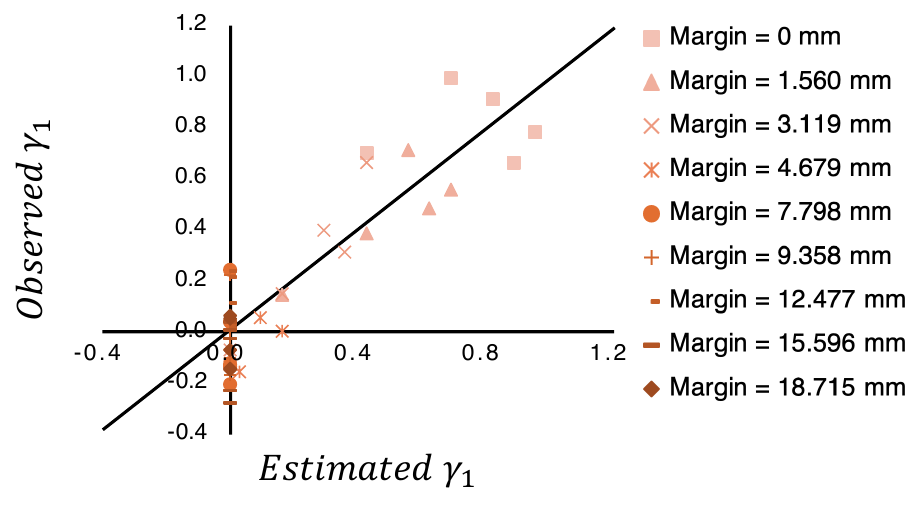}
    \subcaption{Predicted \(\gamma_1\) vs.\ Observed \(\gamma_1\)}
    \label{fig:DualSkewNormal Gamma}
\end{minipage}
\begin{minipage}[b]{0.45\textwidth}
    \centering
    \includegraphics[width=0.95\columnwidth]{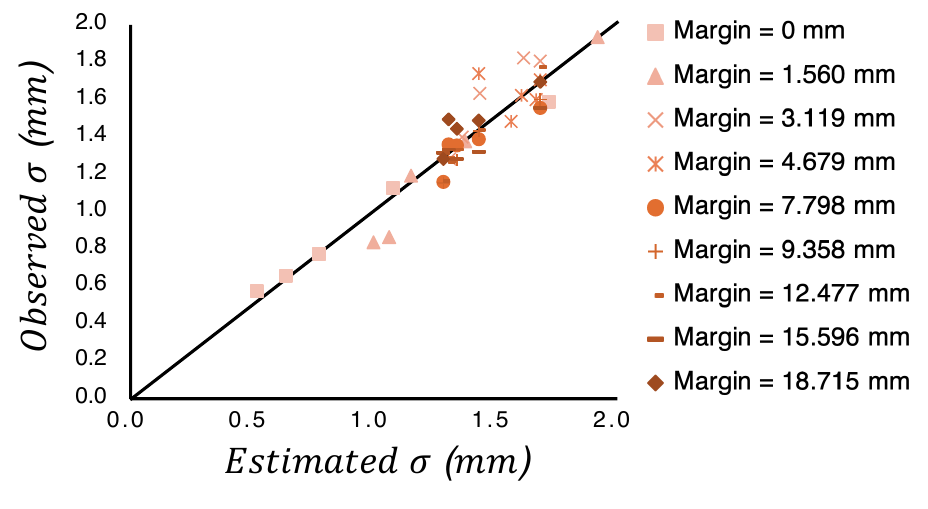}
    \subcaption{Predicted \(\sigma\) vs.\ Observed \(\sigma\)}
    \label{fig:DualSkewNormal Sigma}
\end{minipage}
\begin{minipage}[b]{0.45\textwidth}
    \centering
    \includegraphics[width=0.95\columnwidth]{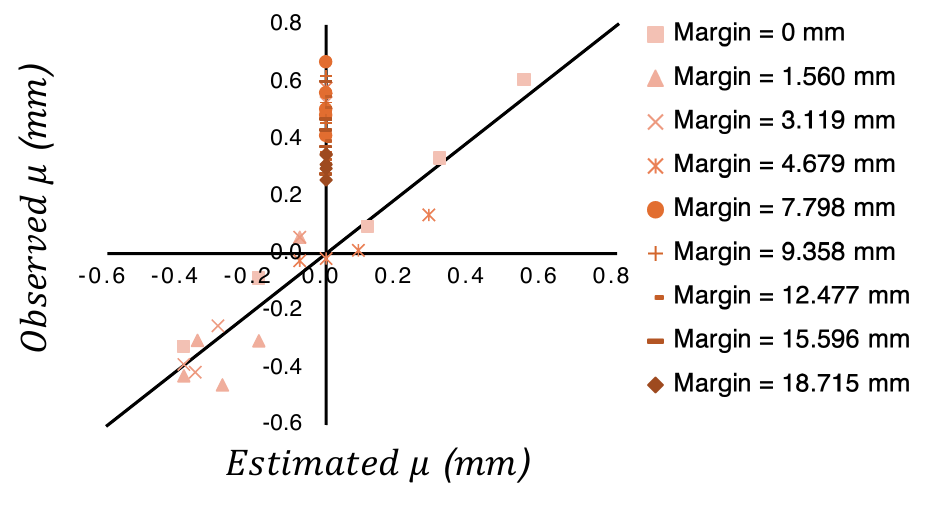}
    \subcaption{Predicted \(\mu\) vs.\ Observed \(\mu\)}
    \label{fig:DualSkewNormal Mu}
\end{minipage}
\begin{minipage}[b]{0.45\textwidth}
    \centering
    \includegraphics[width=0.95\columnwidth]{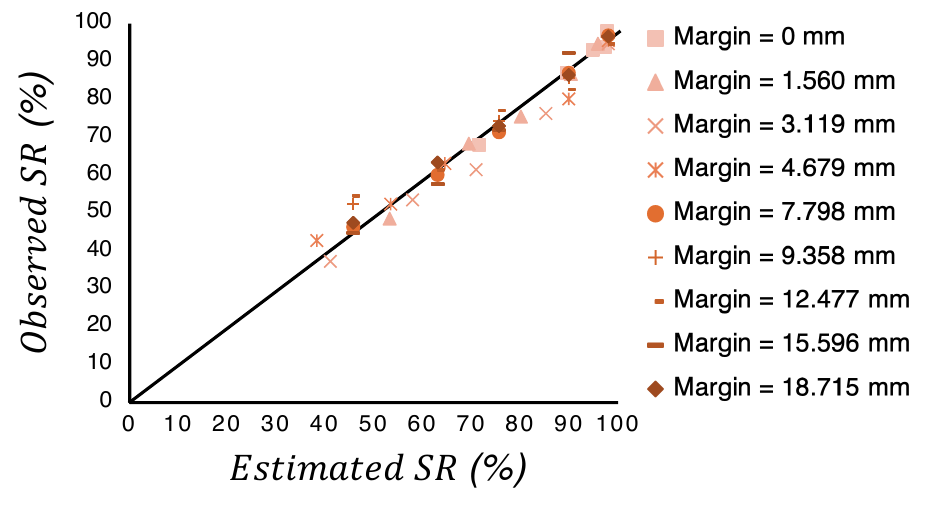}
    \subcaption{Predicted \(\sr\) vs.\ Observed \(\sr\)}
    \label{fig:DualSkewNormal SR}
\end{minipage}
\caption{Results for the Skewed Dual Normal Distribution Model in Experiment 1. Lines indicate perfect prediction. (a) \(\gamma_1\) is well predicted at small \(\margin\), while a slight negative \(\gamma_1\) appears when \(\margin\) is large and we approximated it as 0; this may have been affected by the start position of the task being the right knee of participants. (b) \(\sigma\) is accurately predicted. (c) \(\mu\) is accurately predicted when \HL{\(D_\mathit{edge}<-\frac{c}{d}\); elsewhere, \(\mu\) values were positive.} This was considered to be due to the influence of the start position~\cite{Azenkot2012TouchBehavior,Henze2011LargeExperiment}, finger angle~\cite{Holz2011UnderstandingTouch}, grip~\cite{Lehmann2018HowToHold}, and distance~\cite{Yu2019ERModelVR}. (d) \(\sr\) is accurately predicted overall.}
\Description{Proposed model validation across conditions in Experiment 1. (a) Predicted vs observed skewness $\gamma_1$; (b) predicted vs observed spread $\sigma$ (mm); (c) predicted vs observed mean offset $\mu$ (mm); (d) predicted vs observed $\sr$ (\%). Points cluster along the identity line, showing accurate prediction ($\sr$ $R^2 = .950$), including at small Margins where skew is strongest.}
\end{figure*}

We regressed \(\gamma_1\) using \autoref{Formula:DualSkewNormalEstimateGamma1}, obtaining \(R^2=.789\) (\autoref{fig:DualSkewNormal Gamma}, \autoref{table:Model Regression Ex1}).
As hypothesized, \(|\gamma_1|\) increases (more skew) as the target approaches the edge.
With \( -c/d = 6.40\), the model predicts \(\gamma_1=0\) when the target center is \(\approx 6.40\) mm or farther from the edge; i.e., the tap positions are near a normal distribution.
This result is consistent with the likelihood-ratio test (\autoref{fig:Dedge vs LikelihoodRatio Left}).

We then regressed \(\sigma\) using \autoref{Formula:DualSkewNormalEstimateSigma} by setting $S = W$, with \(c=1.09, d=-0.170\), yielding \(R^2=.882\) (\autoref{fig:DualSkewNormal Sigma}, \autoref{table:Model Regression Ex1}).
\(\sigma\) decreases as the target approaches the screen edge, and the model captures this reduction near the screen edges.
Fitting only the conditions where the model predicts \(\gamma_1=0\) (near-normal) gives \(R^2=.705\), which is better than the Gaussian fit over all data (\(R^2=.437\)).

\HL{Using \(D_\mathit{edge} < -\frac{c}{d}\) data, we regressed \(\mu\) with \autoref{Formula:DualSkewNormalEstimateMu}, obtaining \(R^2=.905\) (\autoref{fig:DualSkewNormal Mu}, \autoref{table:Model Regression Ex1}).
However, for \(D_\mathit{edge} \ge -\frac{c}{d}\), although both our model and the existing model assume that \(\mu\) is approximately 0, we observed positive \(\mu\) values.
This result is probably due to the start position~\cite{Henze2011LargeExperiment}, finger angle~\cite{Azenkot2012TouchBehavior,Holz2011UnderstandingTouch}, grip~\cite{Lehmann2018HowToHold}, and distance~\cite{Yu2019ERModelVR}.}

Combining estimates with Eqs.~\ref{Formula:Delta}–\ref{Formula:Xi} and substituting them into \autoref{Formula:DualSkewSR}, we predicted \(\sr\) and observed unadjusted \(R^2=\HL{.950}\) (\autoref{fig:DualSkewNormal SR}, \autoref{table:Model Regression Ex1}), which was more accurate than the Dual Gaussian model, including at small \(\margin\).
LOOCV analyzes yielded \(R^2\) and \(\mae\) close to the in-sample values, indicating no overfitting and similar generalization to unseen conditions (\autoref{table:Model Regression Ex1}).

\subsubsection{\HL{Machine Learning Models}}
\label{subsubsection:Ex1 ML}
\HL{We also compared our model against several well-established machine-learning baselines that directly predict $\sr$ using $W$ and \(\margin\) as inputs.
The machine learning models used for comparison included Lasso Regression, Random Forest, Support Vector Regression (SVR), and a Multi-layer Perceptron (MLP) Neural Net.
We evaluated these models using two hyperparameter settings: the default values from the scikit-learn library, and values obtained via Bayesian optimization using Optuna library.
In Bayesian optimization, we selected the hyperparameters that minimized the 5-fold cross-validation $\mathit{RMSE}$.}

\HL{For default settings, only Random Forest showed higher \(R^2\) and lower \(\mae\) values than the proposed model (\autoref{table:Model Regression Ex1}).
However, the LOOCV analysis results showed that the $R^2$ were lower than the proposed model.
This suggests that the proposed model yields higher predictive accuracy on unseen data compared to machine learning models with default settings.}

\HL{For optimized settings, SVR and MLP Neural Net showed higher \(R^2\) and lower \(\mae\) than the proposed model (\autoref{table:Model Regression Ex1}).
Even in the LOOCV analysis, the machine learning models achieved higher $R^2$ and lower \(\mae\) than the proposed model.
However, the difference was marginal, at .024 for $R^2$.
Given the ease of interpreting its regression constants, our model is considered more tractable than black-box machine learning models.
For example, in our model, $-c/d$ can be considered as the distance at which tap coordinate skew induced by the screen edge ceases to occur, and \(g\) can be interpreted as a parameter representing the magnitude of \(\margin\)'s influence on \(\sigma\).
This interpretability helps UI designers develop theoretically grounded UI layouts, e.g., they can understand the skewness of the distribution when composing the UI, or they can simply avoid it by using the $-c/d$ threshold.
\HLtwo{In addition, high-accuracy machine learning models are complex simply due to the large number of parameters they require (\autoref{table:Model Regression Ex1}), compared to the 9 parameters ($c$--$l$) in the proposed model.}
Furthermore, the clear internal structure facilitates model extensions such as those for 2D targets~\cite{Usuba2022ER1Dto2D}.}

\subsection{Analyzing Tap Coordinate Distributions}
\begin{figure*}[t]
\centering
\begin{minipage}[b]{0.49\textwidth}
    \centering
    \includegraphics[width=0.95\columnwidth]{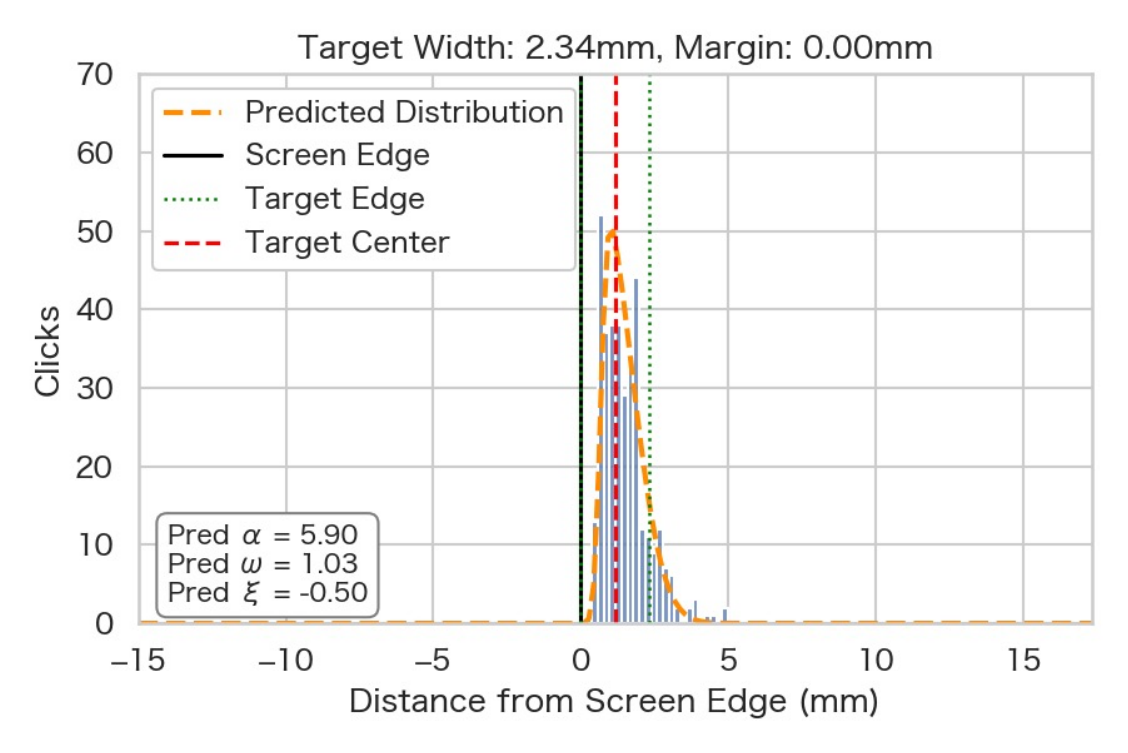}
    \subcaption{No \(\margin\) condition}
    \label{fig:Distribution No Margin}
\end{minipage}
\begin{minipage}[b]{0.49\textwidth}
    \centering
    \includegraphics[width=0.95\columnwidth]{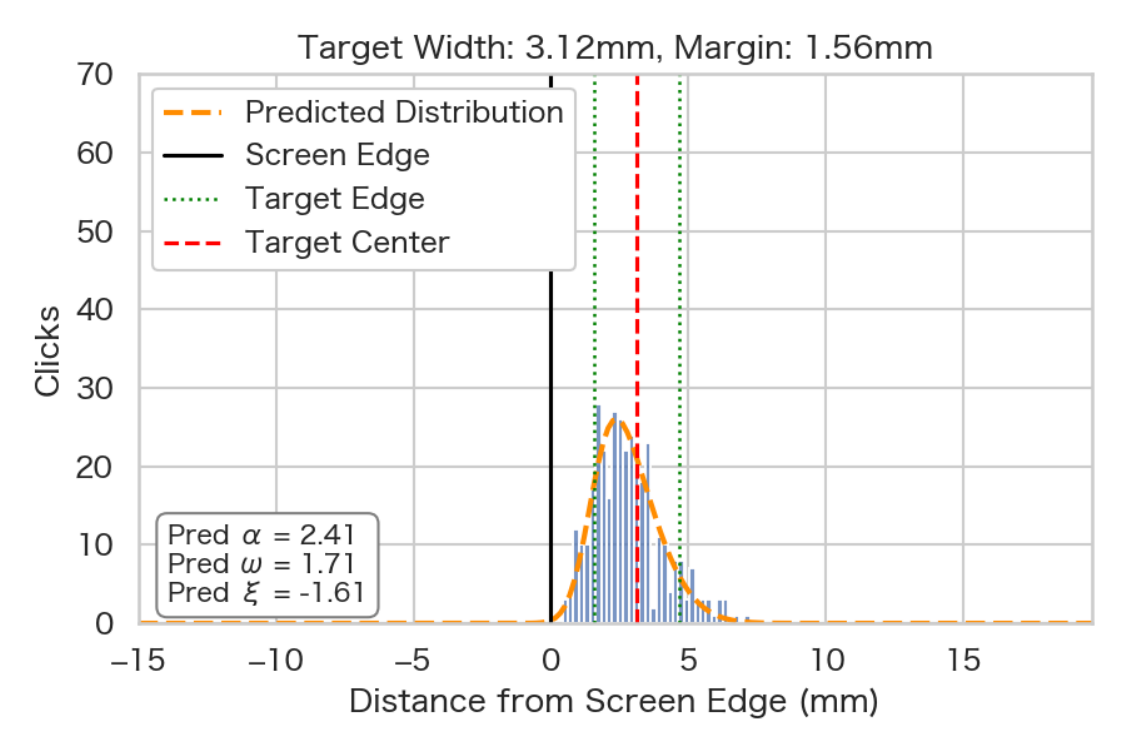}
    \subcaption{Small \(\margin\) condition}
    \label{fig:Distribution Small Margin}
\end{minipage}
\begin{minipage}[b]{0.49\textwidth}
    \centering
    \includegraphics[width=0.95\columnwidth]{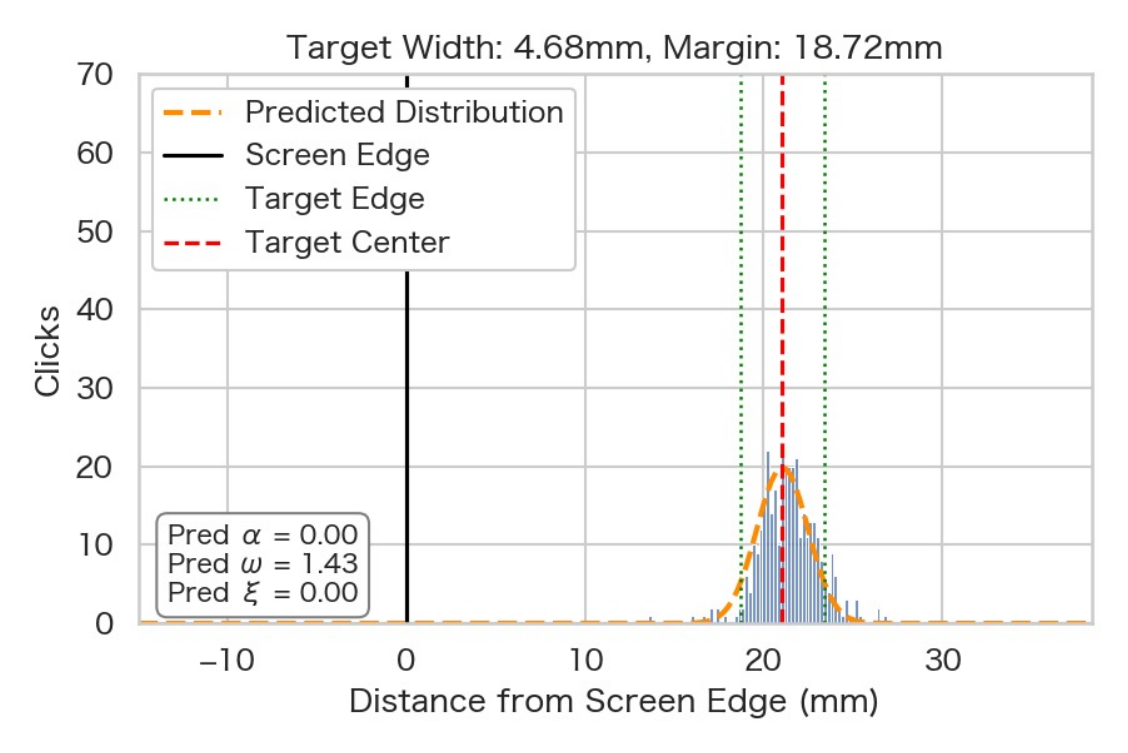}
    \subcaption{Sufficient \(\margin\) condition}
    \label{fig:Distribution Enough Margin}
\end{minipage}
\begin{minipage}[b]{0.49\textwidth}
    \centering
    \includegraphics[width=0.95\columnwidth]{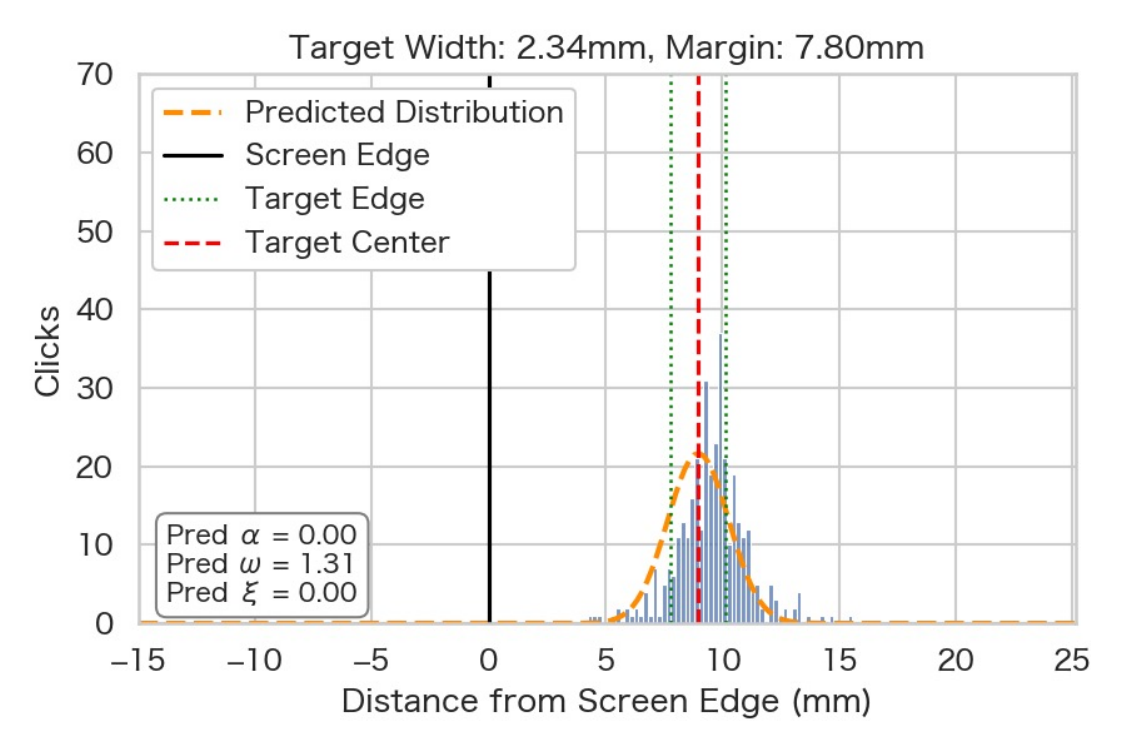}
    \subcaption{Exception condition}
    \label{fig:Distribution Exception}
\end{minipage}
\caption{Observed and predicted tap coordinate distributions in Experiment 1. The proposed model (a) captures extreme skew when the target touches the screen edge and (b) mild skew when near the edge, and (c) returns to the Dual Gaussian Distribution Model when sufficiently far. (d) \HL{In exceptional cases, the distribution shifted rightward on average. This may be attributed to the start position (right knee)~\cite{Yu2019ERModelVR} or the strategy of tapping with an angled finger~\cite{Holz2011UnderstandingTouch}.}}
\Description{Observed tap‑coordinate histograms with model curves in Experiment 1. (a) No‑Margin: strongly skewed toward the edge. (b) Small Margin: mild skew. (c) Sufficient Margin: near‑normal distribution consistent with the Dual Gaussian model. (d) Exception: some conditions showed skew opposite to the edge (peak shifts right), likely due to non‑edge factors like start position, though the model still approximates the shape.}
\label{Fig:TapDistributionX}
\end{figure*}
Analyzing the actual measured tap-coordinate distributions and the distributions assumed by the proposed model suggested that our model can successfully capture the skew induced as the target approaches the screen edge (\autoref{Fig:TapDistributionX}).
The skew predicted by the proposed model successfully modeled both the extreme skew when the screen edge and the target are in contact (\autoref{fig:Distribution No Margin}) and the mild skew when the screen edge and the target are close (\autoref{fig:Distribution Small Margin}).
In addition, when there is sufficient distance, the distribution can be explained by the Dual Gaussian Distribution Model (\autoref{fig:Distribution Enough Margin}).
\HL{However, we also observed exceptional cases where the mean of the distribution shifted to the right of the target (\autoref{fig:Distribution Exception}).
This would stem from non-edge factors, such as the right-side start position of the off-screen task, a strategy of tapping with an angled finger~\cite{Holz2011UnderstandingTouch} and so on.
Still, in most cases, this exception was observed when the target size was large and thus $\sr$ was high.
Due to its minimal impact on $\sr$ estimation, the $\sr$ prediction is sufficiently accurate (\autoref{fig:DualSkewNormal SR}).}

\subsection{Participant Questionnaire}
In open-ended comments, three participants reported deliberately tapping at the boundary between the screen edge and the screen when the target touched the edge.
This supports the strategy of ``tapping together with the edge'' to avoid errors on the opposite side.
Furthermore, the observed tap coordinates were concentrated near the edge.
This suggests that other participants may have implicitly adopted a strategy of tapping near the edge.
Thus, unlike prior reports~\cite{Avrahami2015EdgeTouch,Usuba2023EdgeTarget,Henze2011LargeExperiment}, our observed improvement of \(\sr\) at the edge may stem from flat modern bezels and linear edge–target contact.
Participants also reported diverse strategies, such as tilting the smartphone, tapping with the side of the finger, and pressing the finger—strategies that may explain nonzero \(\mu\) even when far from the edge.

\section{\HL{Results of Experiment 2}}
\subsection{\HL{Outliers}}
\HL{Similar to Experiment 1, after removing the first set, taps falling to the right or left of the target were excluded (58 trials, $0.358\%$).
Then, we removed 70 taps (0.433\%) based on the 3 SD criterion on the $y$-axis~\cite{SOUKOREFF2004751}.
The analyses below utilize the remaining 16,072 trials.
}

\subsection{\HL{Distributional Tests of Tap Coordinates}}
\begin{figure}[t]
\centering
    \includegraphics[width = \columnwidth]{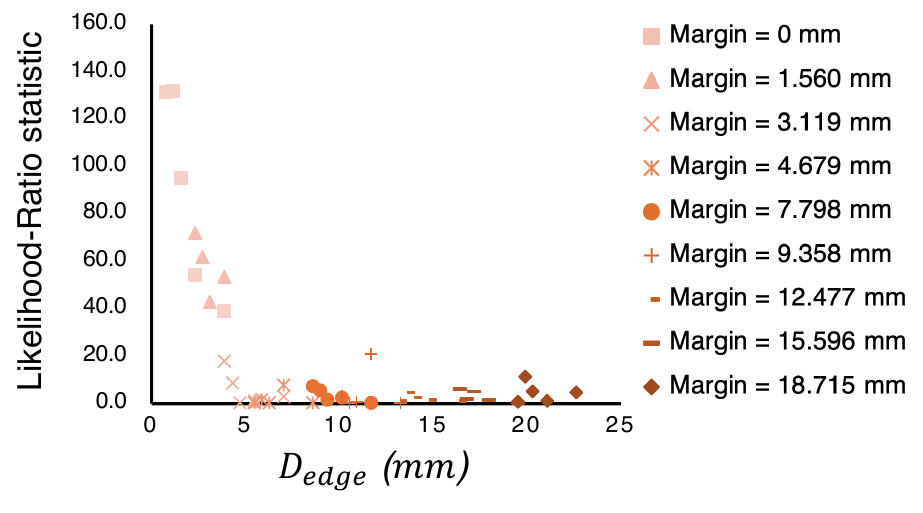}
\caption{\HL{\(D_\mathit{edge}\) vs.\ likelihood-ratio statistic in Experiment 2. Likelihood-ratio tests indicate that the closer the target is to the edge, the more the tap coordinate distribution follows a skew-normal distribution rather than a normal distribution.}}
\Description{Scatter plot of $D_{edge}$ (mm) versus the likelihood‑ratio statistic in Experiment 2 (Bottom Edge). Similar to Experiment 1, the statistic increases as $D_{edge}$ decreases, confirming that vertical tap distributions also become skew-normal near the screen edge.}
\label{fig:Dedge vs LikelihoodRatio Bottom}
\end{figure}

\HL{
The Shapiro–Wilk test results showed that 22 of the 45 conditions (48.9\%) followed a normal distribution.
Likelihood-ratio tests indicated that the tap coordinate distribution fits the skew-normal distribution better as $D_\mathit{edge}$ decreases (\autoref{fig:Dedge vs LikelihoodRatio Bottom}).}

\subsection{\HL{Prediction Accuracy of the Models}}
\begin{table*}[t]
\caption{\HL{Regression constants and accuracy metrics for each model in Experiment 2.} \HLtwo{The number of parameters in machine learning models refers the quantity of parameters necessary for making predictions.} \HL{Although the $R^2$ for the MLP Neural Net was lower than -1, which might appear unnatural, it is a possible outcome\protect\footnotemark[2]. Since MLP Neural Nets strongly require hyperparameter tuning and feature scaling, significantly low $R^2$ values can occur under default settings\protect\footnotemark[3].}}
\centering
\begin{tabular}{c|c|c|c|c|c|c|c|c}
\multicolumn{2}{c|}{} & \multicolumn{5}{c}{Regression Analysis} & \multicolumn{2}{|c}{LOOCV}\\
\hline
Model & Equation & Regression Constants & $R^2$ & $\mae$ & $ \rmse$ & $\mape$ & $R^2$ & $\mae$\\
\hline
Gaussian & $\sigma^2$ (Eq.~\ref{Formula:Bi_sigma}) & \HL{$a = 1.23$, $b = 0.0164$} & \HL{$.370$} & \HL{$0.355$} & \HL{$0.459$} & \HL{$39.1\%$} & \HL{$.315$} & \HL{$0.371$}\\
\cline{2-9}
& $\sr$ (Eq.~\ref{Formula:DualGaussian1D}) & - & \HL{$.699$} &  \HL{$7.51$} & \HL{$11.6$} & \HL{$12.5\%$} & \HL{$.691$} & \HL{$7.61$}\\
\hline
Skewed & $\gamma_1$ (Eq.~\ref{Formula:DualSkewNormalEstimateGamma1}) & \HL{$c = 1.20$, $d = -0.199$} & \HL{$.873$} & \HL{$0.0915$} & \HL{$0.120$} & \HL{$81.6\%$} & \HL{$.867$} & \HL{$0.0944$}\\
\cline{2-9}
& $\sigma$ (Eq.~\ref{Formula:DualSkewNormalEstimateSigma}) & \begin{tabular}{c}\HL{$e = 0.123$, $f = 0.0371$,}\\ \HL{$g = 0.415$, $h = 1.31$,}\\ \HL{$i = 0.0130$}\end{tabular} & \HL{$.871$} & \HL{$0.0675$} & \HL{$0.0915$} & \HL{$5.61\%$} & \HL{$.841$} & \HL{$0.0761$}\\
\cline{2-9}
& $\mu$ (Eq.~\ref{Formula:DualSkewNormalEstimateMu}) & \begin{tabular}{c}\HL{$j = 0.804$, $k = -0.0961$,}\\ \HL{$l = 3.60$}\end{tabular}  & \HL{$.641$} & \HL{$0.124$} & \HL{$0.167$} & \HL{$36.0\%$} & \HL{$.484$} & \HL{$0.153$}\\
\cline{2-9}
& $\sr$ (Eq.~\ref{Formula:DualSkewSR}) & - & \HL{$.953$} & \HL{$3.13$} & \HL{$4.56$} & \HL{$5.24\%$} & \HL{$.947$} & \HL{$3.35$}\\
\hline
\multicolumn{9}{c}{\HL{Machine Learning Models with Default Hyperparameters}}\\
\hline
\HL{Lasso Regression} 
  & \HL{-} & \HLtwo{number of parameters: 3} 
  & \HL{.608} & \HL{10.5} & \HL{13.2} & \HL{16.7\%} 
  & \HL{.558} & \HL{11.1} \\
\hline
\HL{Random Forest} 
  & \HL{-} & \HLtwo{number of parameters: 5,528}
  & \HL{.967} & \HL{2.24} & \HL{3.81} & \HL{3.27\%} 
  & \HL{.799} & \HL{5.82} \\
\hline
\HL{SVR} 
  & \HL{-} & \HLtwo{number of parameters: 46} 
  & \HL{.106} & \HL{16.2} & \HL{19.9} & \HL{28.9\%} 
  & \HL{.065} & \HL{16.7} \\
\hline
\HL{MLP Neural Net} 
  & \HL{-} & \HLtwo{number of parameters: 401} 
  & \HL{-2.49} & \HL{32.1} & \HL{39.4} & \HL{41.4\%} 
  & \HL{-2.08} & \HL{29.9} \\
\hline
\multicolumn{9}{c}{\HL{Machine Learning Models with Tuned Hyperparameters}}\\
\hline
\HL{Lasso Regression} 
  & \HL{-} & \HLtwo{number of parameters: 3} 
  & \HL{.597} & \HL{10.3} & \HL{13.4} & \HL{17.1\%} 
  & \HL{.565} & \HL{10.7} \\
\hline
\HL{Random Forest} 
  & \HL{-} & \HLtwo{number of parameters: 8,171} 
  & \HL{.967} & \HL{2.54} & \HL{3.80} & \HL{3.81\%} 
  & \HL{.806} & \HL{5.67} \\
\hline
\HL{SVR} 
  & \HL{-} & \HLtwo{number of parameters: 44} 
  & \HL{.835} & \HL{3.70} & \HL{8.56} & \HL{4.72\%} 
  & \HL{.753} & \HL{5.59} \\
\hline
\HL{MLP Neural Net} 
  & \HL{-} & \HLtwo{number of parameters: 21,058} 
  & \HL{.999} & \HL{0.180} & \HL{0.307} & \HL{0.295\%} 
  & \HL{.943} & \HL{3.75} \\
\hline
\end{tabular}
\label{table:Model Regression Ex2}
\end{table*}

\subsubsection{\HL{Dual Gaussian Distribution Model}}
\begin{figure*}[t]
\centering
\begin{minipage}[b]{0.45\textwidth}
    \centering
    \includegraphics[width=0.95\columnwidth]{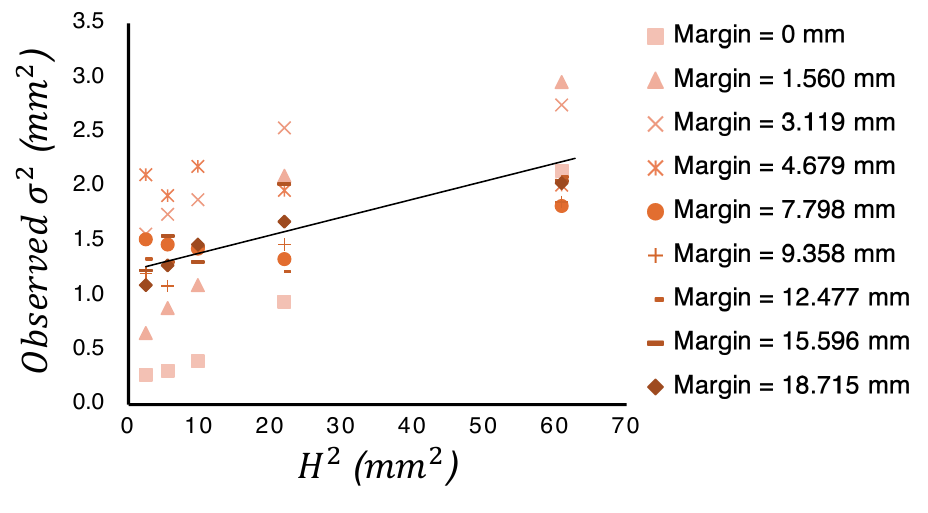}
    \subcaption{\HL{$H^2$ vs.\ $\sigma ^2$}}
    \label{fig:Bottom DualGaussian Sigma}
\end{minipage}
\begin{minipage}[b]{0.45\textwidth}
    \centering
    \includegraphics[width=0.95\columnwidth]{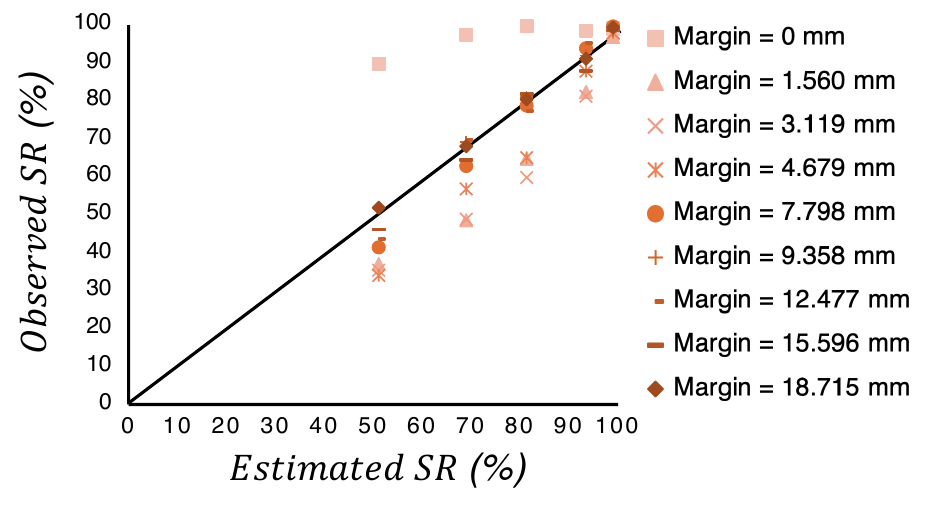}
    \subcaption{\HL{Predicted \(\sr\) vs.\ Observed \(\sr\)}}
    \label{fig:Bottom DualGaussian SR}
\end{minipage}
\caption{\HL{Results for the Dual Gaussian Distribution Model in Experiment 2. Large deviations appear when \(\margin\) is small, indicating the need to model edge distance in \(\sigma\) and \(\sr\). In (a), the line is the regression line; in (b), the line indicates perfect prediction.}}
\Description{Dual Gaussian model diagnostics for Experiment 2. (a) $H^2$ vs observed $\sigma^2$ shows modest fit ($R^2 = .370$) with deviations at small $\margin$s. (b) Predicted $\sr$ vs observed $\sr$ shows degradation in accuracy for small $\margin$ conditions ($R^2 = .699$), confirming the need for edge-aware modeling in the vertical direction.}
\end{figure*}

\HL{We regressed \(\sigma\) of the vertical tap coordinates using \autoref{Formula:Bi_sigma}. 
Fit was modest (\(R^2=.370\); \autoref{fig:Bottom DualGaussian Sigma}, \autoref{table:Model Regression Ex2}), with larger deviations at small \(\margin\).
This trend was consistent with Experiment 1.}

\HL{Predictions of \(\sr\) were also consistent with the results of Experiment 1 (unadjusted \(R^2=.699\); \autoref{fig:Bottom DualGaussian SR}, \autoref{table:Model Regression Ex2}).
\(\sr\) prediction accuracy degraded under conditions with small \(\margin\)s, suggesting the necessity of a model that accounts for the influence of the screen edge, even when the edge is present in the vertical direction.}

\subsubsection{\HL{Proposed Model}}
\begin{figure*}[t]
\centering
\begin{minipage}[b]{0.45\textwidth}
    \centering
    \includegraphics[width=0.95\columnwidth]{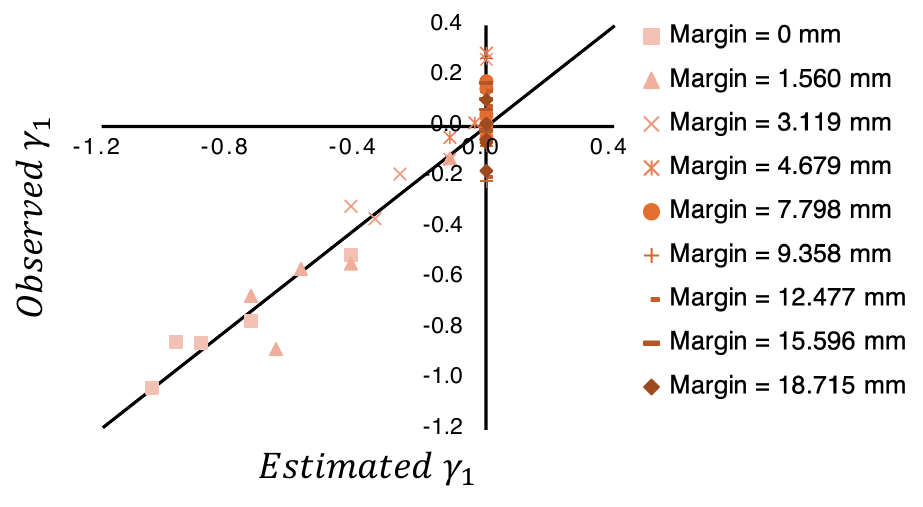}
    \subcaption{Predicted \(\gamma_1\) vs.\ Observed \(\gamma_1\)}
    \label{fig:Bottom DualSkewNormal Gamma}
\end{minipage}
\begin{minipage}[b]{0.45\textwidth}
    \centering
    \includegraphics[width=0.95\columnwidth]{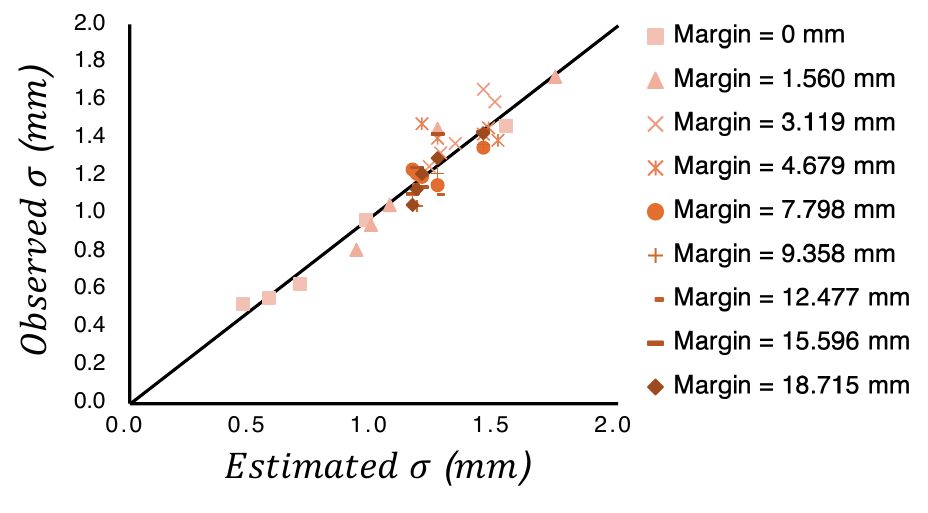}
    \subcaption{Predicted \(\sigma\) vs.\ Observed \(\sigma\)}
    \label{fig:Bottom DualSkewNormal Sigma}
\end{minipage}
\begin{minipage}[b]{0.45\textwidth}
    \centering
    \includegraphics[width=0.95\columnwidth]{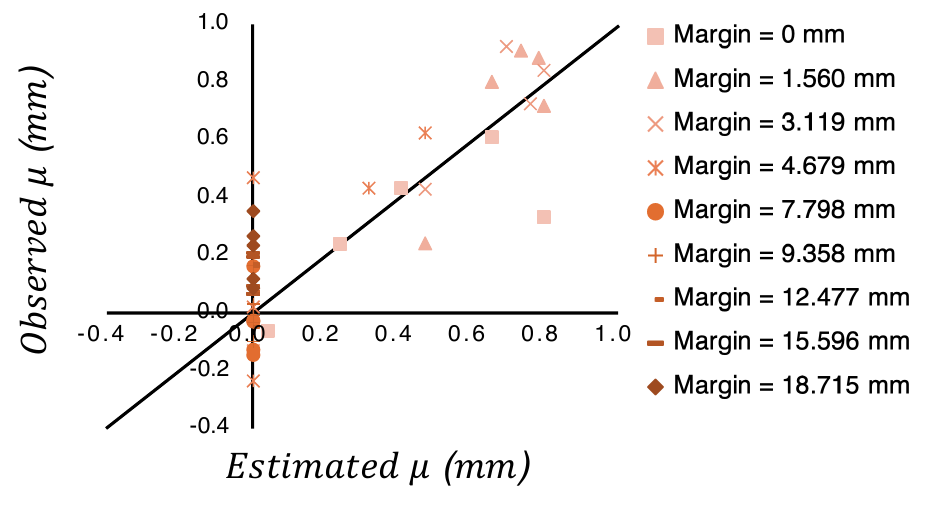}
    \subcaption{Predicted \(\mu\) vs.\ Observed \(\mu\)}
    \label{fig:Bottom DualSkewNormal Mu}
\end{minipage}
\begin{minipage}[b]{0.45\textwidth}
    \centering
    \includegraphics[width=0.95\columnwidth]{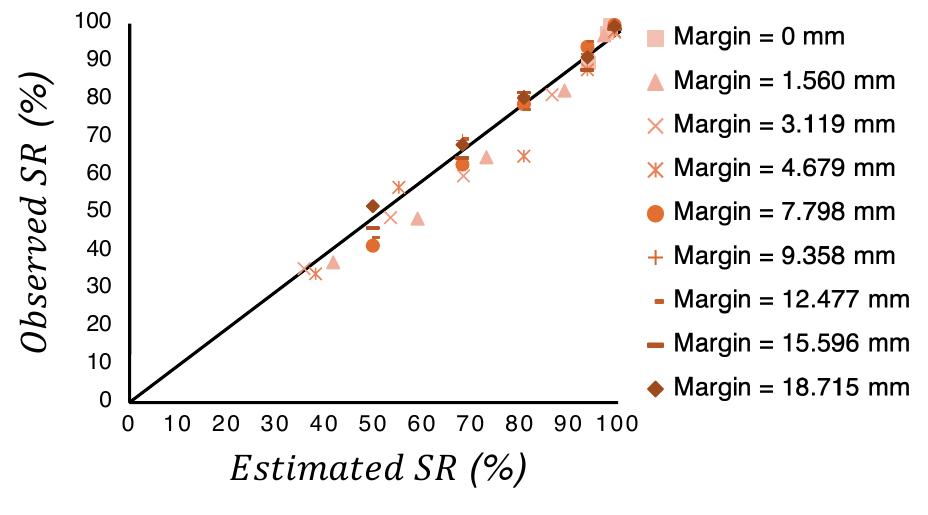}
    \subcaption{Predicted \(\sr\) vs.\ Observed \(\sr\)}
    \label{fig:Bottom DualSkewNormal SR}
\end{minipage}
\caption{\HL{Results for the Skewed Dual Normal Distribution Model in Experiment 2. Lines indicate perfect prediction. (a) \(\gamma_1\) is well predicted. (b) \(\sigma\) is accurately predicted. (c) The prediction accuracy for $\mu$ was lower than in Experiment 1. However, we did not observe large deviations from the model, such as the positive $\mu$ values seen in Experiment 1 when $D_\mathit{edge} \ge -c/d$. The values generally converged to zero for targets far from the screen edge. We attribute this to the start position (right knee), which likely biased the left-edge task rightward but had minimal influence on the bottom-edge task. (d) \(\sr\) is accurately predicted overall.}}
\Description{Proposed model validation across conditions in Experiment 2. (a) Predicted vs observed skewness $\gamma_1$; (b) predicted vs observed spread $\sigma$ (mm); (c) predicted vs observed mean offset $\mu$ (mm); (d) predicted vs observed $\sr$ (\%). The model accurately predicts $\sr$ ($R^2 = .953$) and distribution parameters, with $\mu$ converging to zero for targets far from the edge.}
\end{figure*}

\HL{We predicted \(\gamma_1\) using \autoref{Formula:DualSkewNormalEstimateGamma1}, obtaining \(R^2=.873\) (\autoref{fig:Bottom DualSkewNormal Gamma}, \autoref{table:Model Regression Ex2}).
Consistent with Experiment 1, the $|\gamma_1|$ increased as the target approached the screen edge.
In Experiment 1, $\gamma_1$ was positive because the edge was located in the negative $x$-direction, whereas in Experiment 2, it was negative because the edge was located in the positive $y$-direction (\autoref{fig:EdgeAndTarget Left} and \autoref{fig:EdgeAndTarget Right}).
With \(-c/d = 6.02\), the model predicts \(\gamma_1=0\) when the target center is \(\approx 6.02\) mm or farther from the edge.
This result is consistent with Experiment 1 and the likelihood-ratio test (\autoref{fig:Dedge vs LikelihoodRatio Bottom}).
}

\HL{We predicted \(\sigma\) using \autoref{Formula:DualSkewNormalEstimateSigma} by setting $S = H$, with \(c=1.20, d=-0.199\), yielding \(R^2=.871\) (\autoref{fig:Bottom DualSkewNormal Sigma}, \autoref{table:Model Regression Ex2}).
This result showed the improvement in prediction accuracy for $\sigma$ by our model, and it is consistent with Experiment 1.
}

\HL{Using \(D_\mathit{edge} < -\frac{c}{d}\) data, we regressed \(\mu\) with \autoref{Formula:DualSkewNormalEstimateMu}, obtaining \(R^2=.641\) (\autoref{fig:Bottom DualSkewNormal Mu}, \autoref{table:Model Regression Ex1}).
The prediction accuracy was lower than Experiment 1 ($R^2=.905$).
We attributed this decrease in prediction accuracy to the influence of gravity.
For example, participants may have adopted different strategies in response to gravity (e.g., some tapping higher anticipating gravity, others tapping lower following gravity), potentially resulting in unstable \(\mu\).
On the other hand, deviations of $\mu$ from zero for $D_\mathit{edge} \ge -\frac{c}{d}$ were minimal.
We attribute this to the start position (right knee): while it biased tap coordinates rightward in the left-edge, its influence was weaker in the bottom-edge.
}

\HL{
We predicted \(\sr\) and observed unadjusted \(R^2=\HL{.953}\) (\autoref{fig:Bottom DualSkewNormal SR}, \autoref{table:Model Regression Ex2}).
LOOCV analyses yielded \(R^2\) and \(\mae\) close to the in-sample values, indicating no overfitting (\autoref{table:Model Regression Ex2}).
These results demonstrate that the proposed model can consistently predict $\sr$ accurately even for targets near the bottom edge.}

\subsubsection{\HL{Machine Learning Models}}
\HL{As in Experiment 1, we compared the proposed model with machine learning models (\autoref{table:Model Regression Ex2}).
With default settings, only Random Forest showed a higher $R^2$ than the proposed model, but the proposed model achieved a higher LOOCV $R^2$.
Therefore, the proposed model has higher predictive accuracy for unseen conditions.
With optimized parameters, Random Forest and the MLP Neural Net showed higher $R^2$ values than the proposed model.
However, unlike Experiment 1, the proposed model demonstrated a higher LOOCV $R^2$.
This suggests that for targets near the bottom edge, the proposed model can predict $\sr$s more accurately than machine learning models.
Furthermore, the advantages regarding the interpretability and ease of extension remain consistent here.}

\subsection{\HL{Analyzing Tap Coordinate Distributions}} 
\begin{figure*}[t]
\centering
\begin{minipage}[b]{0.49\textwidth}
    \centering
    \includegraphics[width=0.95\columnwidth]{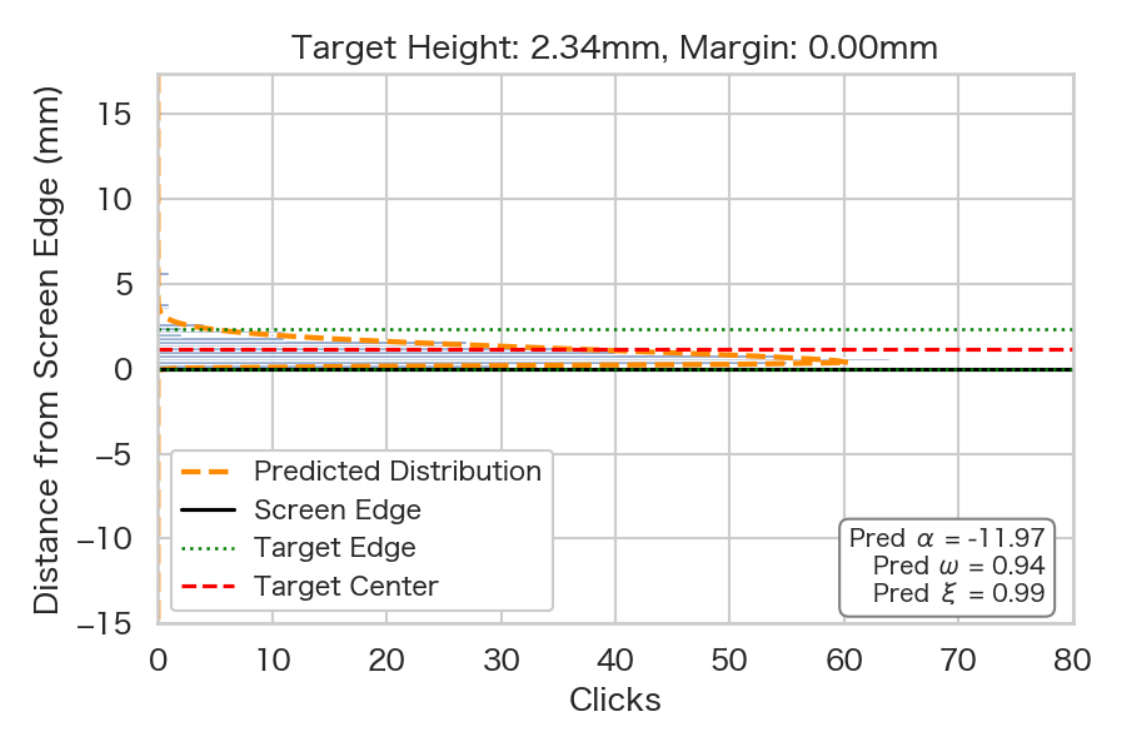}
    \subcaption{\HL{No \(\margin\) condition}}
    \label{fig:Bottom Distribution No Margin}
\end{minipage}
\begin{minipage}[b]{0.49\textwidth}
    \centering
    \includegraphics[width=0.95\columnwidth]{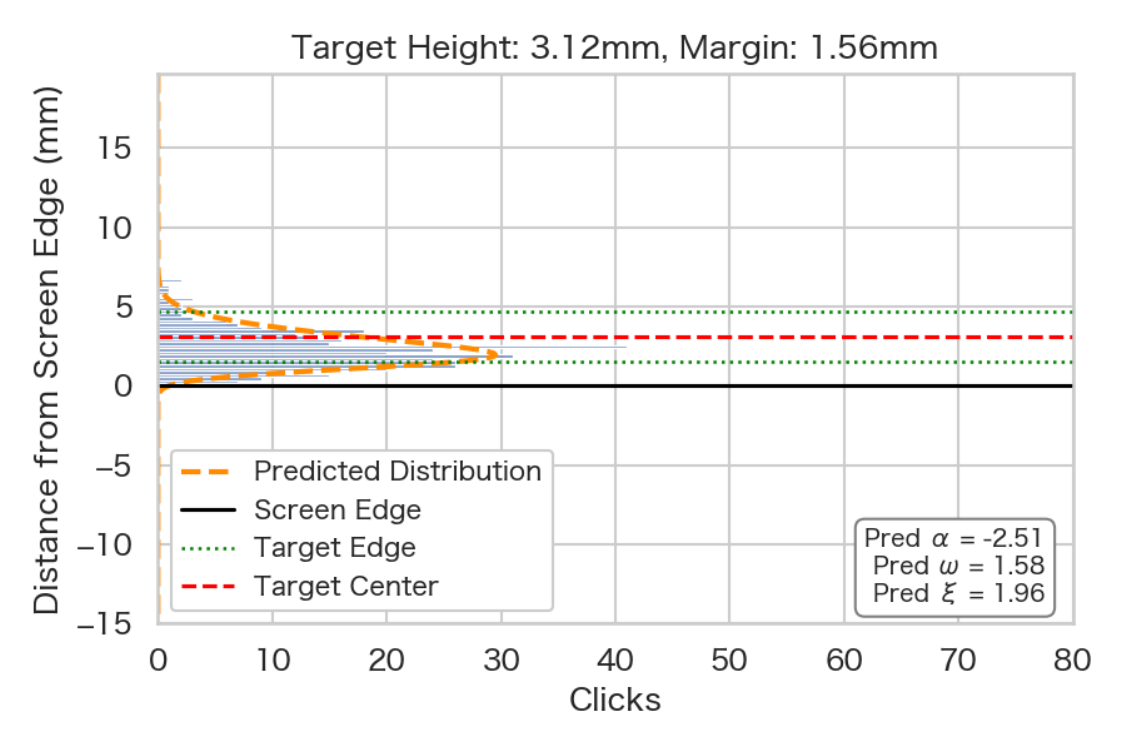}
    \subcaption{\HL{Small \(\margin\) condition}}
    \label{fig:Bottom Distribution Small Margin}
\end{minipage}
\begin{minipage}[b]{0.49\textwidth}
    \centering
    \includegraphics[width=0.95\columnwidth]{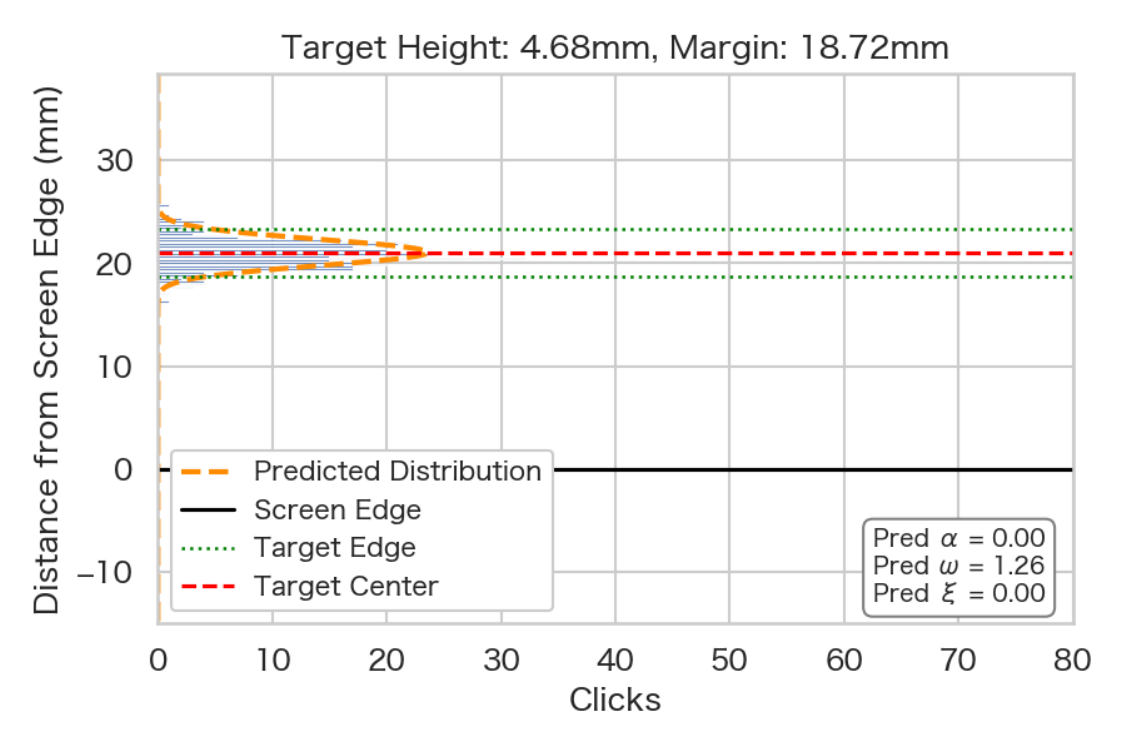}
    \subcaption{\HL{Sufficient \(\margin\) condition}}
    \label{fig:Bottom Distribution Enough Margin}
\end{minipage}
\begin{minipage}[b]{0.49\textwidth}
    \centering
    \includegraphics[width=0.95\columnwidth]{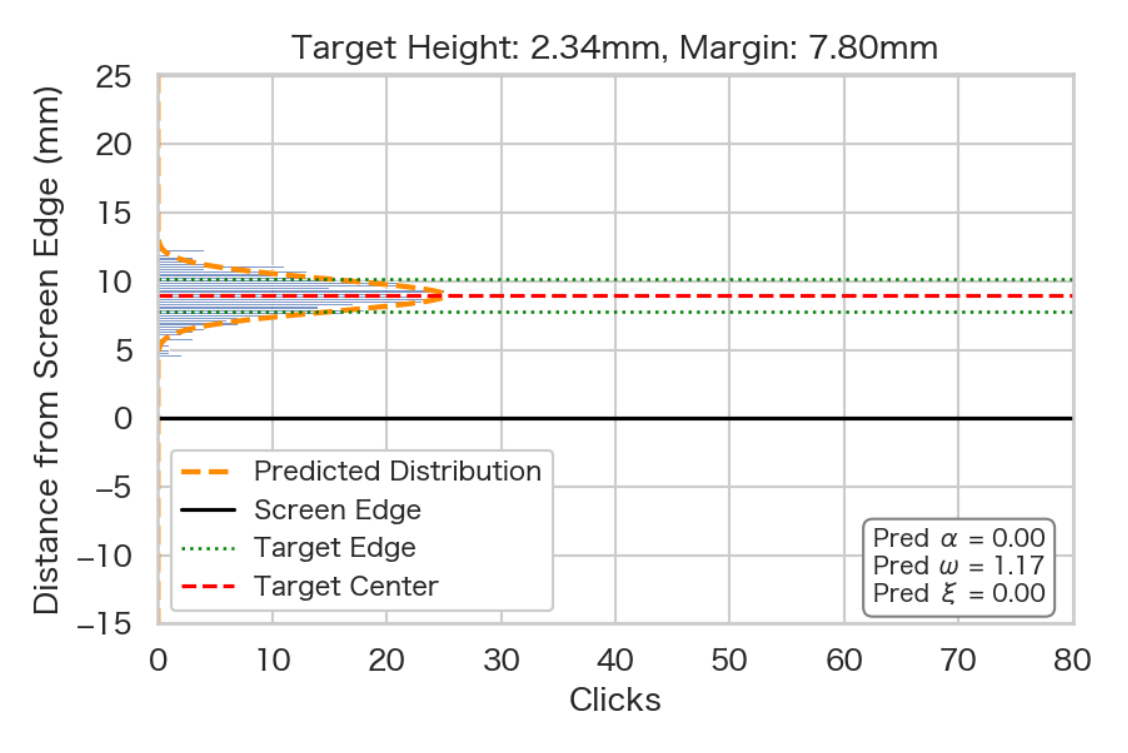}
    \subcaption{\HL{No rightward shift observed in Experiment 2}}
    \label{fig:Bottom Distribution Exception}
\end{minipage}
\caption{\HL{Observed and predicted tap coordinate distributions in Experiment 2. The proposed model (a) captures extreme skew when the target touches the screen edge and (b) mild skew when near the edge, and (c) returns to the Dual Gaussian Distribution Model when sufficiently far. (d) In contrast to Experiment 1, the exception of an average rightward shift in the distribution was not observed in Experiment 2. This is likely because the rightward shift induced by the start position (right knee) had minimal influence in the bottom-edge task.}}
\Description{Observed and predicted tap coordinate distributions in Experiment 2. (a) No‑Margin: extreme skew. (b) Small Margin: mild skew. (c) Sufficient Margin: near-normal distribution. (d) Unlike Experiment 1, no significant reverse shift was observed, likely because the start position (right knee) had minimal influence on the vertical task.}
\label{Fig:TapDistributionY}
\end{figure*}

\HL{Analysis of the tap coordinate distributions suggested that, consistent with Experiment 1, the proposed model can appropriately model the skew of the vertical distribution (Figs. \ref{fig:Bottom Distribution No Margin}--\ref{fig:Bottom Distribution Enough Margin}).
Furthermore, we no longer observed a shift of the entire distribution (\autoref{fig:Bottom Distribution Exception}), such as the rightward shift seen in Experiment 1 (\autoref{fig:Distribution Exception}).
We attribute this to the start position's rightward bias not translating into a vertical shift.}

\subsection{\HL{Participant Questionnaire}}
\HL{Similar to Experiment 1, three participants reported ``deliberately tapping the target together with the screen edge'' when the target touched the edge.
Additionally, six participants reported deliberately tapping above the target.
This is likely due to the discrepancy between the user's perceived aiming point and the actual registered tap location~\cite{Holz2010FingerShift,Holz2011UnderstandingTouch}.}

\section{Discussion}
\subsection{Findings}
\HL{The overarching objective of this work is to provide an interpretable mathematical model that predicts average $\sr$ across the entire screen, including near edges, based solely on UI layout parameters.}
We demonstrated that \HLtwo{1D} tap coordinate distributions near the edge were well modeled by our model.
This extends model coverage to areas previously treated as exceptions.
When the target center is sufficiently far from the edge, the distributions are near-normal, and our model reverts to the Dual Gaussian Distribution Model.

We also provide insights into user strategies.
Prior work has suggested that users avoid edges, thereby degrading performance~\cite{Henze2011LargeExperiment,Avrahami2015EdgeTouch,Usuba2023EdgeTarget}, while our results showed that $\sr$s improved when the target touched the edge.
This difference likely arises from device and target geometry: prior studies used protruding frames~\cite{Avrahami2015EdgeTouch} and circular targets~\cite{Henze2011LargeExperiment,Usuba2023EdgeTarget}, making simultaneous edge–target tapping difficult.
On the other hand, our device (Google Pixel 6a) has a flat bezel, and we used rectangular targets.
Consequently, the strategy of ``tapping the target together with the edge'' became rational to reduce errors on the side opposite the bezel.
In addition, the physical edge may serve as a spatial anchor (placeholder effect~\cite{Adam2006MovingFarther,Bradi2009ModulatingFitts,Pratt2007VisualLayout}).

Finally, our results suggest quantitative guidelines.
Based on the regression constants, when the target center is \(\gtrsim 6.40\) mm from the left edge \HL{and \(\gtrsim 6.02\) mm from the bottom edge}, tap coordinates are approximately normal.
If UI designers wish to avoid edge effects, maintaining \HL{these margins} is recommended.
However, as we observed an improved \(\sr\) when the target touches the edge, placing targets at the edge may be reasonable in some cases.

\subsection{Implications}
Our work offers the following implications for future researchers.
\begin{itemize}
  \item Our \HL{two} experiments reveal that tap distributions skew for targets within $\approx 6$ mm, limiting the applicability of normality-assuming models to non-edge regions.
  \item To predict \(\sr\) across the screen, our model provides a \HL{unified one-dimensional formulation} that accounts for skew \HL{and reduces to the Dual Gaussian Distribution Model when edge-induced skew disappears}.
  \item Modern flat bezels and linear edge-target contact enable users to tap targets together with the edge. Models and studies should consider this possibility.
  \item \HL{While tuned machine learning models can achieve comparable or slightly higher $R^2$, our model offers advantages over black-box approaches due to its analytic structure. Interpretable parameters like $-c/d$ and a transparent computation process make our model easier to interpret, communicate to designers, and extend.}
\end{itemize}

\subsection{Limitations and Future Work}
We restricted tasks to 1D pointing for tractability, but since real UIs are 2D, further validation for rectangular targets is needed~\cite{Bi2016DualGaussian,Usuba2022ER1Dto2D}.
\HL{For Dual Gaussian model, Usuba et al.\ demonstrated that the 2D $\sr$ can be predicted by the product of two 1D $\sr$ models estimated from horizontal- and vertical- tasks~\cite{Usuba2022ER1Dto2D}.
This suggests that our 1D model might also serve as a building block for an edge-aware 2D model (\autoref{Formula:Dual Skew 2D}).}
\begin{equation}
\begin{split}
\sr &= P\left(-\frac{W}{2} \leq X \leq \frac{W}{2}\right) \times P\left(-\frac{H}{2} \leq Y \leq \frac{H}{2}\right) \\
&= \int_{-\frac{W}{2}}^{\frac{W}{2}}f(x)dx \times \int_{-\frac{H}{2}}^{\frac{H}{2}}f(y)dy
\end{split}
\label{Formula:Dual Skew 2D}
\end{equation}
\HL{Here, $\int_{a_1}^{a_2}f(a)da$ represents the probability mass of the skew-normal distribution within the interval $[a_1, a_2]$.}
\HL{Verifying whether such a factorization remains valid is included in our future work.}

We used dominant index-finger off-screen pointing by right-handed users.
\HL{Thus, another validation is needed for one-handed thumb input~\cite{Perry2008Thumb} and non-dominant hand usage.}
We only tested the left \HL{and bottom} edges, so additional work should examine the right/top edges, and corners.
\HL{Furthermore, this study validated only the accuracy of the $\sr$ prediction.
Thus, a detailed validation of the model's derivation itself is left for future work.}

Since the bezel shape and size may affect strategy, testing devices other than Google Pixel 6a is important.
\HL{Also, understanding the influence of cases with different shapes is also important.
As a hypothesis, because the case creates an untappable area near the screen edge, we may have to re-estimate the model parameters for each smartphone case.}

\HL{The proposed model estimates average $\sr$ based solely on UI layout parameters, since the objective of this study was to develop a mathematical model as a tool for UI designers.
However, a user's subjective speed-accuracy bias can also influence tap accuracy~\cite{yamanaka23CHI}, and incorporating its influence is our future work.
\HLtwo{For example, the model by Wobbrock et al.\ predicts success rates based on the speed–accuracy tradeoff, where attempting to shorten movement time degrades the success rate~\cite{Wobbrock2008_ERModel}.
Incorporating the interaction between screen edge effects, movement time, and success rates into the model is a promising future direction.}
Still, in research on the Dual Gaussian Distribution Model, the consensus has been to estimate $\sr$ without restricting $\mt$s~\cite{Bi2016DualGaussian,Yamanaka2020Rethinking,Yamanaka2024ISS}.
}

Currently, our model sets $\gamma_1=0$ and $\mu=0$ when the target is sufficiently far from the screen edge.
However, experimental results indicated some deviations, such as an average rightward shift of tap coordinates observed even when the target was far from the left edge.
Modeling these factors in detail to further improve prediction accuracy is one future direction.

Lastly, integrating our model into UI design tools (e.g. Tappy~\cite{usuba24arxivTappy,Yamanaka24arXivFigmaTappy}, Tap Analyzer~\cite{LIFULL24tapAnalyzer}) is promising.
\HL{For example, it may be useful to quantitatively visualizing the decrease in $\sr$ as a target approaches the screen edge due to scrolling.
Furthermore, unlike machine learning models, our model offers the advantage of high explainability, as it allows for visualizing how changes in the distribution lead to changes in $\sr$.}

\section{Conclusion}
In this work, we demonstrated \HL{via two 1D smartphone experiments (left/bottom edges)} that our Skewed Dual Normal Distribution Model outperforms the Dual Gaussian Distribution Model in predicting success rates near screen edges.
Our model captures the skew of tap distributions near edges and smoothly reverts to the Dual Gaussian form when the target center is about $>6~mm$ away from the edge.
\HL{Unlike machine-learning baselines which achieve similar accuracy but lack transparency, our analytic model exposes interpretable parameters (e.g., skew thresholds) valuable for UI layout reasoning and theory-based design.}
Our work advances understanding of behavior near edges and extends the Dual Gaussian Distribution Model to edge-adjacent targets, contributing to the development of theory-based UI design tools.

\bibliographystyle{ACM-Reference-Format}
\bibliography{sample-base}
\end{document}